\documentclass[useAMS,usenatbib]{mn2e}
\usepackage{graphicx}
\usepackage{amsmath}
\newcommand{\hangpar}{\noindent\hangindent.1in}
\newcommand{\teff}[1]{$T_{\rm eff}$}
\newcommand{\vsini}[1]{$v\cdot\sin(i)$}

\title[Stratification and Isotopic Separation in CP Stars]
{Stratification and Isotope Separation
in CP Stars\thanks{Based on observations
obtained at the European Southern
Observatory, Paranal and La Silla, Chile
(ESO programmes 65.L-0316(a), 68.D-0254(A),
076.D-0169(A) and 081.D-0498(A).}}
\author[C. R. Cowley, S. Hubrig,
and J. F. Gonz\'{a}lez]
{C. R. Cowley${^1}$,
\thanks{E-mail: cowley@umich.ed}
 S. Hubrig${^2}$,
and J. F. Gonz\'{a}lez${^3}$   \\
$^{1}$Department of Astronomy, University of Michigan,
   Ann Arbor, MI 48109-1042, USA\\
$^{2}$European Southern Observatory, Casilla 19001,
Santiago 19, Chile\\
${^3}$Complejo Astron\'{o}mico El Leoncito, Casilla 467, 5400
San Juan, Argentina
}
\begin{document}

\date{Accepted . Received ; in original form }

\pagerange{\pageref{firstpage}--\pageref{lastpage}} \pubyear{2008}

\maketitle

\label{firstpage}

\begin{abstract}
We investigate the elemental and isotopic stratification
in the atmospheres of selected chemically peculiar (CP) stars
of the upper main sequence.
Reconfiguration of the UVES spectrograph in 2004 has
made it possible to examine all three lines of the
Ca {\sc ii} infrared triplet.  Much
of the material analyzed was obtained in 2008.

We support the claim of
Ryabchikova, Kochukhov \& Bagnulo (RKB) that
the calcium isotopes have distinct stratification
profiles for the stars 10 Aql, HR 1217, and HD 122970,
with the heavy isotope
concentrated toward the higher layers.
Better observations
are needed to learn the extent to which $^{40}$Ca
dominates in the deepest layers of all or most CP stars
that show the presence of $^{48}$Ca.
There is little evidence for $^{40}$Ca in the spectra
of some HgMn stars, and
the infrared triplet
in the magnetic star HD 101065 is well
fit by pure $^{48}$Ca.   In HR 5623 (HD 133792)
and HD 217522
it is likely that the heavy isotope dominates,
though models are possible where this is not the case.

While elemental stratification is surely needed
in many cases, we
point out the importance of including
adjustments in the assumed $T_{\rm eff}$
and $\log(g)$ values, in attempts to model stratification.
We recommend emphasis
on profiles of the strongest lines, where the influence
of stratification is most evident.

Isotopic mixtures, involving the 4 stable calcium
nuclides with masses between 40 and 48 are plausible,
but are not emphasized.

\end{abstract}

\begin{keywords}
stars:atmospheres--stars:chemically peculiar--stars: magnetic fields
--stars:abundances
--stars:individual: HR 1217   
--stars:individual: HR 1800   
--stars:individual: HD 101065
--stars:individual: HD 122970
--stars:individual: HR 5623   
--stars:individual: HR 7143   
--stars:individual: 10 Aql    
--stars:individual: HR 7245   
--stars:individual: HD 217522
\end{keywords}

\section{Rationale and introduction}

The current study was undertaken to solidify our knowledge
of chemical and isotopic stratification of calcium in
chemically peculiar (CP) stars of the upper main sequence.
We hope such knowledge will lead to an improved
understanding of the complex physical processes taking
place in the atmospheres of these stars.

Previous work (cf. Cowley and Hubrig 2005, henceforth
Paper I)
has demonstrated clearly the presence of
rare isotopes of calcium in stars as different as the
field HZB star Feige 86 ($T_{\rm eff} = 16430$K) and
Przybylski's star (HD 101065, $T_{\rm eff} = 6600$K).

Lines of the Ca {\sc ii} infrared triplet (IRT) have
easily measurable isotope shifts, very nearly 0.20~\AA\, between
$^{48}$Ca and $^{40}$Ca for all three lines
(N\"{o}rtersh\"{a}user, et al. 1998).  The
large shifts arise
because of the unusual
nature of the 3d orbitals of
the ground term of the IRT; they have collapsed below
the 4p subshell.
Other Ca {\sc ii} lines show far
smaller isotopic shifts, of the order of milliangstroms.

In a few cases, e.g. the HgMn star
HR 7143 (Castelli and Hubrig 2004), the isotope-sensitive
lines appear both symmetrical, and shifted entirely to
the wavelengths of the rare isotope, $^{48}$Ca.
This isotope comprises only some 0.2\% of terrestrial
calcium.


Ryabchikova (2005) and her coworkers find that in
roAp stars
the cores of the profiles indicate $^{48}$Ca, but
the wings are arguably produced by the common isotope
$^{40}$Ca.

If only the cores of the isotope-sensitive lines are
shifted, the observations may be reproduced by a model
with a thin layer of the rare heavy calcium isotope.
In this case, the relative amount of the exotic species,
in terms of a column density above optical depth unity,
could be quite small--far smaller than if the bulk of
the line absorption were due to $^{48}$Ca.  It is important
to know which, if either,
of these scenarios is dominant.

We have examined several line profiles in some detail for
seven stars.  The following discussion is based on several
possible models, with and without stratification.  In the
former case, we computed profiles based on both elemental
and elemental plus isotopic stratification.
Automated as well as trial and error methods were used.
Details of all models and techniques considered would not
be appropriate here.  We present an eclectic resum\'{e}.
Specific details are available on request from CRC.

\section{Elemental stratification}

It is generally accepted that the outer layers of CP
stars are chemically differentiated from their bulk composition.
The mechanism responsible for this separation (Michaud 1970)
is capable of producing differentiation {\it within} the
photosphere, or line-forming regions of these stars.  Such
separation is now widely referred to as {\it stratification}
(cf. Dworetsky 2004).  Early indications of the need for
vertical, chemical or density structures that depart from a
classical one-dimensional, chemically homogeneous
 atmospheric structure were described
by Babcock (1958), and analyzed in some detail in a series
of papers by Babel (cf. Babel 1994).

The most striking indications of stratification are
in the cores of the Ca {\sc ii} resonance lines, particularly
the K-line.
Babel (1992) proposed a wind model with an
abundance profile that reproduced the
sharp, deep cores of the H- and K-lines
(see his Fig. 4).

Cowley, Hubrig \& Kamp
(2006) presented a short atlas of K-line cores
in CP and normal stars.  They
also showed (cf. their \S6)
that an ad hoc modification of the
temperature distribution would also give cores
similar to those illustrated in their paper.
A sharp drop in the {\it overall} atmospheric density
in a chemically homogeneous atmosphere would also
produce sharp K-line cores.  However, work by
Ryabchikova and her collaborators
(e.g. Ryabchikova, Kochukhov \& Bagnulo 2008,
henceforth, RKB)
show different stratification patterns for
different elements, that  exclude models
with chemical homogeneity.

LeBlanc \& Monin (2004) discuss calculations
somewhat similar to those of Babel, though
without a wind.  They also obtain stratification
profiles similar to those which reproduce
observations.

\subsection{Modeling elemental stratification}



There are no models of stellar atmospheres
with elemental stratification built in from first
principles, and most researchers
have used an empirical approach.
While Babel's work focused on the strong Ca II K-line,
subsequent stratification studies have
employ various lines, of more
than one ionization stage.
Strong and weak lines were used, including
the IRT lines.


Kochukhov, et al. (2006, henceforth KTR)
derive stratification parameters by a ``regularized
solution of the vertical inversion problem'' (VIP).
They apply the technique to the magnetic
CP star HR 5623 (HD 133792).
The sophistication of the method notwithstanding,
VIP lacked a significant generality in practice.
KTR first {\it fixed} $T_{\rm eff}$,
$\log(g)$, $\xi_t =0$, and $v\cdot\sin(i) = 0$,
and used them to derive calculated spectra.
These fundamental parameters also affect the basic
observed minus calculated values used to obtain the
stratification profiles.
{\it Thus, an error in
$T_{\rm eff}$ or $\log(g)$ could be reflected in
erroneous stratification parameters.}
In principle, the difference between observed and calculated
spectrum should consider all relevant parameters {\it including}
those specifically describing the stratification.

We discuss the model for HR 5623 below, and conclude
that the model parameters are
not easy to fix for this star.

Ryabchikova, Leone,
and Kochukhov (2005) and subsequent papers by these
workers describe the code DDAFIT, which is based on a
limited set of 4 parameters describing the stratification.

Both DDAFIT and the VIP method  derive stratification profiles
from a comparison of the observed and calculated
spectra.  If applied to any single line profile,
DDAFIT would be similar to our method
(cf. $g(x)$ and $g_{48}(x)$ below).
DDAFIT does
assume a sharp boundary between domains with different
isotopic compositions, while our functions smooth over these
boundaries.  DDAFIT automatically adjusts its parameters
to achieve an optimum fit with the help of a
Levenberg-Marquardt routine (Kochukhov 2007).


In two previous papers (Cowley, et al. 2007, henceforth, Paper II,
Cowley \& Hubrig 2008, henceforth Paper III),
we used stratification profiles
for calcium based on
an analytical function, $g(x)$,
and four parameters,
$a$, $b$, $d$, and in an obvious notation, the abundance
${\rm Ca}/N_{\rm tot}$ in the
deepest photosphere:

\begin{equation}
g(x) = b + (1-b)\left[\frac{1}{2}\pm
{\frac{1}{2}}{\rm erf}(\sqrt{(a|x+d|^2)}\right].
\end{equation}

\noindent Here, $x = \log{(\tau_{5000})}$; the
abundance at any depth, $x$, is
$g(x)\cdot {\rm Ca}/N_{\rm tot}$.
The negative sign is taken for $x < -d$ (see
Fig.~\ref{fig:strats}).

We used Atlas 9 models, as implemented and described by
Sbordone et al. (2004) to obtain $T(\tau_{5000})$.
Pressures, opacities, and line profiles were obtained
with Michigan software described in previous publications.

We have used both a trial-and-error method and a least
squares minimization based on the (downhill) simplex routine
UMPOL from the IMSL (1998) library.

\subsection{Strong vs. weak lines as stratification indicators}

In this work, we have tried to avoid weak lines, preferring
the strong lines of Ca {\sc ii}, either
the K-line, or lines of the IRT.
In some stars, the resonance lines of Sr {\sc ii} show the
characteristics of stratification (Paper III).

Strong lines have
several advantages over weaker ones.  First, the effects
of stratification are much larger, as may be
seen by comparing Figs. 7 and 9 of Paper III
for the strong Ca {\sc ii} lines $\lambda\lambda$3933 and
8542 with the Fig. 14, where we were at some pains
to show the effect of stratification on the
subordinate Sr {\sc ii} line, $\lambda$4162.

Much of the discrepancy
between observed and calculated weak and
intermediate-strength lines
is in the line depths.  Line depths have
subtle dependences on many factors, instrumental,
model dependent ($T_{\rm eff}$, microturbulence,
$v\cdot\sin(i)$, etc.),
and atomic
($gf$-values, damping, hfs).  One may readily get
a fit for any individual line depth by adjusting one
or more of these parameters.
By contrast, the observed profiles of stronger
lines are less subject to perturbations by noise,
blends, and the instrumental profiles.  We know
of no reasonable adjustment of
parameters that could reconcile the observed, anomalous cores
with those calculated for the strong Ca {\sc ii} K-line
using a classical model.
However,
the generally-accepted stratification models can
fit these strong-line profiles.


\section{Isotopic stratification}

It has been known for decades that isotopic anomalies
occur in the atmospheres of CP stars (see the review
by Cowley, Hubrig, and Castelli, 2008).

These anomalies, like the chemical peculiarities,
are not believed to reflect the bulk compositions
of the stars.
While it is assumed that
the isotopic separations are caused by the same kinds
of forces
that give rise to the overall chemical
peculiarities, detailed explanations of the anomalies
remain to be worked out.

In a few cases, the isotope separation mechanism has
been so efficient that the material remaining in the
line-forming regions is virtually isotopically pure.
A canonical case has been mercury in $\chi$ Lupi,
which is some 99\%-pure $^{204}$Hg.  Proffitt, et al.
(1999) give references and an extensive discussion.
Various efforts have been made to establish
stratification of mercury.  The even-A isotopes of
mercury are well separated in wavelength in certain
sharp-lined spectra (cf. Woolf and Lambert 1999), but
we know of no convincing studies showing different formation
strata for mercury isotopes.

In the case of stars showing anomalously strong lines
of $^3$He, Bohlender (2005) has concluded that the lighter
isotope is in layers above those with the normal isotope
$^4$He.  He finds that the Stark widths are systematically
smaller for the lighter isotope, indicating that it is
formed in higher regions of the atmospheres with lower
gas pressures.

\subsection{Modeling isotopic stratification}

Ryabchikova (2005) and her coworkers
used models with the
heaviest isotopes ($^{48}$Ca and/or $^{46}$Ca)
concentrated above
$\log(\tau_{5000}) \approx -1.3$.  The common
$^{40}$Ca dominates the deeper layers. In the
deepest layers, the Ca/H-ratio can exceed the solar value
by more than two orders of magnitude (HR 5105, HR 7575).

\begin{figure}
\includegraphics[width=54mm,height=84mm,angle=270]{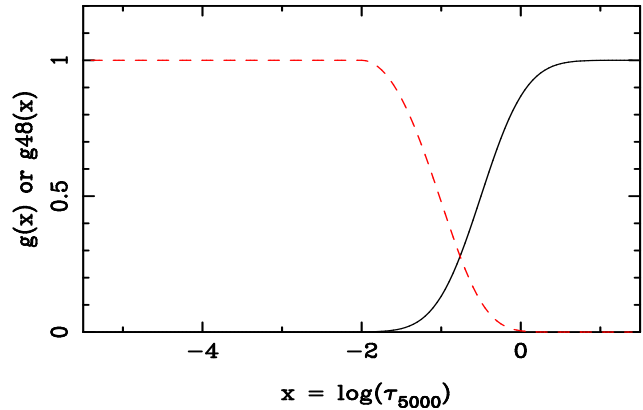}
 \caption{Stratification functions $g(x)$ (solid)
and $g_{48}(x)$
(dashed), where $x = \tau_{5000}$.
Relevant parameters for the plots are in Tab~\ref{tab:10AB}
for $\lambda$8542.
Note the minimum of $g(x)$ is not zero, but determined by
the parameter $b$.  Both $b$ and $g_{48}(x)$ are
set to zero for values of $x$
smaller than $10^{-8}$.
\label{fig:strats}}
\end{figure}

We avoid an abrupt transition  in the isotopic
mix by introducing a second
function, $g_{48}(x)$, to simulate a layer rich in the
heavy isotope (or isotopes).  Again, $x = \log(\tau_{5000})$.
This function places the center of a cloud
of exotic calcium  at a depth
$x = -d'$.  The function
$g_{48}(x)$ is defined to be unity for
$x = \pm q-d'$.  On either side of this domain, the
function declines rapidly to zero.  By an appropriate choice
of parameters, the upper boundary of the cloud can be
put above the highest layers in the atmosphere, as
illustrated in Fig.~\ref{fig:strats}.

\begin{equation}
g_{48} = 1.0 - {\rm erf}(a'x_q^2),
\end{equation}
\noindent with
\begin{equation}
x_q = |x+d'| - q,
\end{equation}

Fig.~\ref{fig:strats} shows a case with the heavy
isotope is effectively restricted to layers above
$\log(\tau_{5000})$ ca.  $-$0.8.


The simplex fits tend to
push the centroid of the cloud very high in the
model.  This tendency had already been noted by
Ryabchikova (2005).
Additional study of this problem requires a hyperextended
atmosphere including non-LTE, which we leave for
future work.

\subsection{Column densities\label{sec:colden}}

In a stratified atmosphere there is no single
${\rm Ca}/N_{\rm tot}$ ratio.  As a substitute
one may consider integral column densities,
for some
``equivalent'' column length, $H$.
We adopt the following, somewhat arbitrary
definition.

\begin{equation}
<N_{\rm Ca}H> =
\int_{\tau_\lambda({\rm min})}^{\tau_{\lambda({\rm max})}}
N_{\rm Ca}\exp(-1.5\cdot \tau_\lambda)
\frac{d\tau_\lambda}{\kappa_\lambda}.
\label{eq:NH}
\end{equation}

\noindent The integrals are taken from the smallest optical
depth of our models to the largest, or from
$\log(\tau_{5000} = -5.4$ to 1.4.

The $N_{\rm Ca}$-values are calculated with the help of
the model atmosphere, and the relevant stratification
profile, $g(x)$ or $g_{48}(x)$.  With this
definition, we can show that very different column
densities of $^{48}$Ca arise in the models with and
without isotopic stratification.

A related column density is that of all massive particles.
In an obvious notation,
\begin{equation}
<N_{\rm tot}H> =
\int_{\tau_\lambda({\rm min})}^{\tau_{\lambda({\rm max})}}
 \frac{P_g-P_e}{kT}\exp(-1.5\cdot \tau_\lambda)
\frac{d\tau_\lambda}{\kappa_\lambda}.
\label{eq:NHtot}
\end{equation}

From these relations, we may make rough
intercomparisons of elemental abundances in stratified
and unstratified atmospheres.

\subsection{Variable log($gf$)'s}

In order to allow for variable relative abundances of
individual calcium isotopes, we often adjusted the
log($gf$) values for the IRT lines independently.
Since the line absorption coefficient
involves the {\it product} of the abundance and
oscillator strength, increasing the $f$- or
$gf$-value for a given line is equivalent to increasing
the abundance for that line.  The elemental and atomic
data input to the calculation includes a
{\it provisional} (note the prime) ratio
${\rm Ca'}/N_{\rm tot}$, where $N_{\rm tot} = (P_g-P_e)/(kT)$,
(massive particles).  When a good line fit is
achieved, the provisional ${\rm Ca'}/N_{\rm tot}$ is
the optimum abundance ratio {\it for that particular
line, provided the assumed $gf$-value is also the
adopted one}.  If the $gf$-value differs from that
adopted, the abundance that corresponds to a line fit
must be modified.  Logarithmically,

\begin{eqnarray}
\lefteqn{\log({\rm Ca}/N_{\rm tot})_{\rm adopted} =
 \log({\rm Ca'}/N_{\rm tot})_{\rm provisional}} \nonumber  \\
&  & +\log(gf)_{\rm used} - \log(gf)_{\rm adopted}
\end{eqnarray}

\noindent In this work we have only assumed the presence
of $^{40}$Ca and $^{48}$Ca.  In some cases, better fits to
the observations could have been obtained by including
intermediate isotopes, but this has not been done for
the present study.

For reference,
Tab~\ref{tab:gf} lists values of log($gf$) from
VALD, Mel\'{e}ndez, Bautista, and Badnell (2007, MBB), and
Brage et al. (1993, Tab. IV, Col. 3).
Convenience motivated our adoption of
VALD values, although they are probably less accurate
than those of MBB or Brage et al.

\begin{table}
 \centering
  \caption{Recent log($gf$)-values for Ca {\sc ii} IRT}
  \begin{tabular}{l l l l}    \hline
  $\lambda$ [\AA] & VALD  & Brage et al.  & MBB \\ \hline
  8498   &$-$1.416 &$-$1.369     &$-$1.366    \\
  8542   &$-$0.463 &$-$0.410     &$-$0.412    \\
  8662   &$-$0.723 &$-$0.679     &$-$0.675    \\
\end{tabular}
\label{tab:gf}
\end{table}

\section{The observational basis for separation
of $^{40}$Ca, $^{48}$Ca, and perhaps other Ca isotopes}
\label{sec:obs}

\begin{figure}
\includegraphics[width=55mm,height=84mm,angle=270]{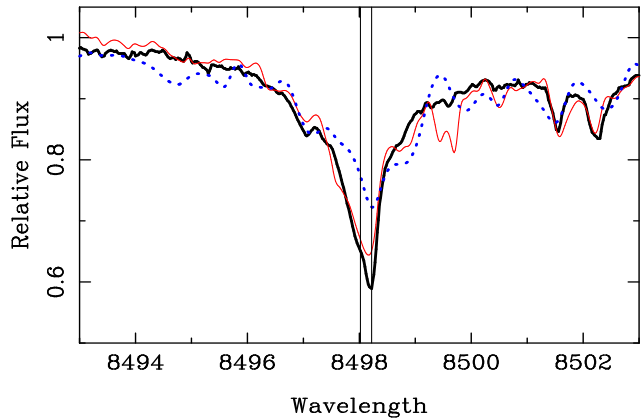}
\caption{Three reductions of the line $\lambda$8498 of
the IRT.  All are based on UVES spectra taken on 8 October
2001 of 10 Aql.  The vertical lines are at the positions
for $^{40}$Ca (8498.02), and $^{48}$Ca (8498.22).  In order
of increasing depth at the latter wavelength, the reductions
were made (1) for the ESO archive,
\newline http://archive.eso.org/eso/eso\_archive\_adp.html
UVES Pipeline 3.9.0
(dotted: blue in online version,
gray in b/w), (2) for SH in 2006 with UVES Pipeline 2.9.0
and mildly Fourier filtered
(thin solid line: red online version, darker gray
in b/w), and
(3) reduced by FG using IRAF in 2008 (thick: black).
Neither (1) nor (3)
were Fourier filtered, but all three spectra were rectified
as described in Paper III \S3.
\label{fig:3A}}
\end{figure}

Most of the relevant observational material for
isotopic stratification of heavy calcium has been obtained
with the UV-Visual Echelle Spectrograph (UVES) at UT2 of the
VLT.  The instrument and spectra have been described
elsewhere (cf. Castelli and Hubrig 2004).
Observers may start from the same raw observations, and
get spectra that can be significantly different
because of the way the material is processed.  This
seems to be critically true in the region of the Ca {\sc ii} IRT,
and is illustrated in Fig.~\ref{fig:3A}.

Epochs of the spectra illustrated here are given in the
figure captions.  Many were obtained in August of 2008,
and reduced especially for the present study by FG.  The
remainder were reduced with pipeline programs current
for their epoch.

A relatively small number of stars are suitable for the
study of isotopic separation in calcium.  First, the large isotopic
shifts occur only for the IRT lines.  Second, a very small
fraction of CP stars show the largest shifts, as may be seen
in Figs. 1 and 2 of Paper II.  Tab.~\ref{tab:bigshifts} lists
the CP stars with the largest averaged shifts
from Paper II.  For these stars, the average measured shift
of $\lambda\lambda$8498 and 8662 (as well as 8542, when available)
is $\ge 0.15$~\AA.  The seven roAp stars were all included in the
study by RKB.  Note that HgMn stars are among
those with the largest shifts.

\subsection{Wavelengths, isotopes, and stratification models}

The plots in Papers I and II show unequivocally that
the wavelength shifts of all three lines of the IRT
are highly correlated.  However, the shifts are
significantly different in the spectra of
(magnetic) CP2 stars (Preston 1974).
Measurements of published and
recently obtained spectra show that shifts for the $\lambda$8662
are 0.06~\AA\, larger on the average, than for
$\lambda$8498.  Average shifts for $\lambda$8542 are similar
to those for $\lambda$8498, though in important individual
cases (10 Aql, $\gamma$ Equ), the shifts increase from the
shortest to the longest wavelength line.

We estimated (Paper II) that any individual
wavelength measurement
might be uncertain by up to 0.05~\AA.
These uncertainties could be due to a variety of
causes, such as proximity to order gaps, the asymmetry
of the line profiles, or to blends.
Whatever their cause, shifts of the order of 0.06~\AA\,
are easily measurable, and readily detected in our
figures.


At present, we admit that significant
differences in the wavelength shifts of IRT lines
exist in individual spectra.  Their cause has not
yet been resolved.


\begin{table}
 \centering
  \caption{CP stars with large IRT wavelength shifts}
  \begin{tabular}{r l l c}    \hline
  HD Number & Other  & Type  & Average Shift \\
            &designation&    &          \\ \hline
  24712  &HR 1217  & roAp     &0.16 \\
  65949  &         & mercury  &0.15  \\
 101065  &Przybylski's&roAp   &0.20  \\
 122970  &         & roAp     &0.16  \\
 133792  &HR 5623  & roAp     &0.18  \\
 134214  &         & roAp     &0.18  \\
 175640  &HR 7143  & HgMn     &0.20  \\
 176232  &10 Aql   & roAp     &0.18  \\
 178065  &HR 7245  & HgMn     &0.16  \\
 217522  &         & roAp     &0.20 \\ \hline
\end{tabular}
\label{tab:bigshifts}
\end{table}

\subsection{Isotopic stratification: different core and
wing shifts}

Traditional detections of isotopic mixtures or anomalies
in stellar atmospheres have been based primarily on wavelength
shifts.  RKB present plots showing that the mean wavelengths of
the $wings$ and cores of IRT  lines show different shifts.

These findings are illustrated
in their figures 6 and 7, which include four of the
stars of Tab.~\ref{tab:bigshifts}.
Their Figure 6 is of the $\lambda$8498-line of the sharp-lined
spectrum of 10 Aql;  figure 6b shows that a calculation
assuming a 50-50\% mix of $^{40}$Ca and $^{46}$Ca+$^{48}$Ca
have a wing that is deeper than the observed red wing.  The
core has a minimum at 8498.20~\AA, which would
correspond to pure $^{48}$Ca.  They conclude the heavy isotope(s)
dominate only in the uppermost layers.

\begin{figure}
\includegraphics[width=55mm,height=84mm,angle=270]{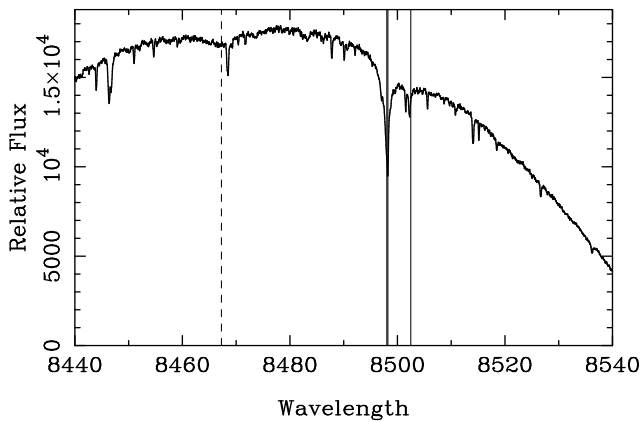}
\caption{Single order UVES spectrum (8 October 2001) of
10 Aql near $\lambda$8498.
The dashed line marks the position of P17.
The thick vertical line
(actually two virtually unresolved lines)
marks the wavelengths of $^{40}$Ca
and $^{48}$Ca.  The centroid of  P16 is shown by the vertical line
furthest to the right, near a blend of two Si I lines.}
\label{fig:fgraw}
\end{figure}

These workers note the difficulty in establishing an accurate
observational profile in Section 2 of their paper.
We entirely concur (cf. Fig.~\ref{fig:3A}).  Their
procedure replaces a
poor, observed
P16 profile by a theoretical one.  However, in order
to remove the flawed profile, it is necessary to
disentangle it from the Ca {\sc ii} line, and this is
not straightforward.  One can see this from
Fig.~\ref{fig:fgraw}, which shows the unrectified profile
of a single order for 10 Aql, as reduced by FG.

We first discuss cases where the evidence for isotopic
stratification is strong, and then turn to stars for
which the indication of such separation is marginal
or absent.

\section{HR 1217 (HD 24712)}

HR 1217 is the best case that we have examined for
isotopic separation.
RKB's Figure 7 shows observational and calculated
fits for the intermediate-strength line, $\lambda$8662
as well as $\lambda$8498.
Their best fit is
shown to be the one with high layers dominated by
heavy calcium--isotopic stratification.
In Paper II, we reported shifts of
0.17 and 0.15 (respectively) for these two line cores.
Our measurements were from a UVESPOP archive
spectrum (Bagnulo,
et al. 2003), not from the same instrument as used
by RKB.

\begin{figure}
\includegraphics[width=85mm,height=84mm,angle=0]{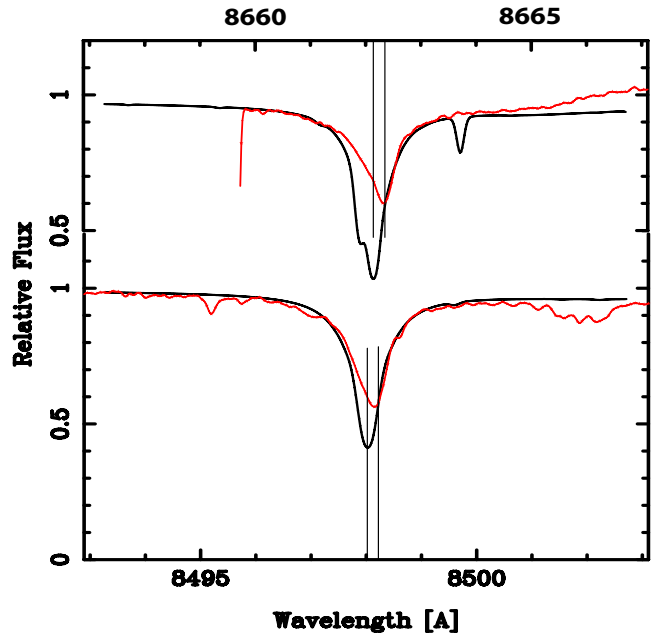}
 \caption{Calculated (black) and observed spectra
(gray with dots, red in online version) of IRT lines
$\lambda$8662 (above), and 8498 (below) in HR 1217.
The UVES spectrum (14 March 2001) is from the UVESPOP archive.
The calculations are intended only to show the match
of the line wings with an assumption of pure $^{40}$Ca.
The two vertical lines mark the rest wavelengths for
$^{40}$Ca (left) and $^{48}$Ca (right).  The observed
red wing of $\lambda$8662 is strongly affected by the order
break, and we have moved it to vertically.  Clearly
lines centered at the position of the vertical lines
to the right ($^{48}$Ca) would {\it not} fit the
observed wings.
\label{fig:wing9862}}
\end{figure}

We confirm from the UVESPOP spectrum that
both lines are
readily fit with the heavier isotope,
$^{48}$Ca, providing the shifted
core.   Fig.~\ref{fig:wing9862} is based on the UVESPOP
spectrum, and shows that the {\it wings}  of both the
$\lambda$8498 and $\lambda$8662 lines agree with a profile
calculated with $^{40}$Ca only.


\section{The Ca {\sc ii} IRT in 10 Aql}

The 10 Aql model used below has $T_{\rm eff} = 7650$K,
$\log(g) = 4.0$, and with solar abundances replaced
by appropriate averages (e.g. for neutrals and ions)
from Ryabchikova, et al. (2000).

New measurements of the wavelengths of the cores of the IRT
lines were obtained from UVES spectra obtained on 4 August
2008.  We obtained shifts of +0.17, +0.19, and +0.20~\AA\,,
for $\lambda\lambda$8498, 8542, and 8662, respectively.
The measured shifts reported in Paper II were +0.14, and
+0.22\AA\, for $\lambda\lambda$8498 and 8662.  The differences
are consistent with our error estimates for measurements
of asymmetrical features.

\subsection{The $\lambda$8498 line\label{sec:9810}}

\begin{figure}
\includegraphics[width=84mm,height=84mm,angle=0]{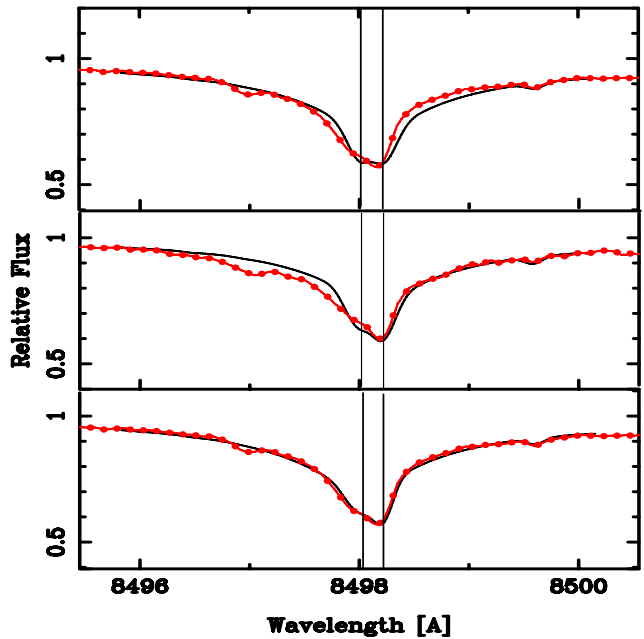}
 \caption{The $\lambda$8498 line in 10 Aql (UVES
4 August 2008).  Observed spectra are dark gray (red online)
with points, calculations in black.  Vertical lines mark the
wavelengths for $^{40}$Ca and $^{48}$Ca.
The upper and middle plots are for a uniform
{\it isotopic} mix through atmosphere,
but {\it elemental} stratification.
Lower plot assumes isotopic {\it and} elemental stratification.
Relevant parameters for the fit are in Tab~\ref{tab:10AB}.
\label{fig:8498c}}
\end{figure}

The $\lambda$8498 line is the weakest of the IRT.
Fig.~\ref{fig:8498c} (upper) shows the result of an
automatic (simplex) calculation
of the $\lambda$8498 region, using the stratification
and abundance
parameters for $g(x)$ given in the second column of
Tab.~\ref{tab:10AB}.
For the upper and middle plots, we assumed
no variation in isotopic abundances with depth.
In the deepest layers of the atmosphere, where $g(x)$ is
unity, the value log(Ca$'/N_{\rm tot})$ {\it is} the relative
abundance provided the oscillator strength is the
one accepted.  If we accept the VALD value, -1.416,
then $\rm \log(^{48}Ca/N_{tot}) = -3.75$.  The abundance for
the more common isotope is smaller by 0.04 dex, because
a smaller oscillator strength was required to fit
the part of the feature dominated by $^{40}$Ca.  Adding
the isotopes, we get $-$3.47 for
$\rm \log[(^{40}Ca+^{48}Ca)/N_{tot}]$.

The center plot of Fig.~\ref{fig:8498c} results from
trial and error (t\&e) adjustments of the parameters to
get a better fit near the position of the $^{48}$Ca
core at 8498.22~\AA.  It is arguable whether an improvement
has been achieved, but the total calcium in the deeper
layers, log[($^{40}$Ca+$^{48}$Ca)$/N_{\rm tot}]$ = $-$4.07.

The parameters in the 4th column produced the fit shown
in the bottom plot of Fig.~\ref{fig:8498c}.  This was
also a trial and error fit, but adjusted from a simplex
solution.  The automated result pushed the $^{48}$Ca
cloud so high that only a ``sliver'' of a region remained
with the heavy isotope.  Under these conditions,
we did not believe the column density calculation was
realistic.  Note that the $d$-parameter of $g(x)$ is
quite small, and this requires a relatively high
Ca/$N_{\rm tot}$ ratio in the deepest layers.


\begin{table}
 \centering
  \caption{Parameters for 10 Aql fits with (iso-strat)
  and without (uniform) isotopic
  stratification.  The trial and error (t\&e) solution
  was also made with a uniform isotopic ratio.
  Abundances are in the rows labeled
  log(Ca$'/N_{\rm tot})$.
  These values are multiplied by $g(x)$ or $g_{48}(x)$
  respectively in the profile calculation.}
\setlength{\tabcolsep}{4pt}
\begin{tabular}{l r r r r r}    \hline
   &\multicolumn{3}{c}{$\lambda$8498}&\multicolumn{2}{c}{$\lambda$8542} \\
Param.            &uniform & t\&e    & iso-strat&  uniform   &iso-strat \\  \hline
$a$               &4.28    & 4.0     &  2.23    & 4.289      &2.0    \\
$\log(b)$         &$-$4.72 &$-$4.19  &$-6.00$   &$-$5.133    &$-$6.30 \\
$d$               & 0.4376 & 0.4     & 0.110    & 0.1936     &0.0  \\
log(Ca$'/N_{\rm tot})$&$-$3.75 &$-$4.19  &$-$3.25   &$-$3.03     &$-$3.30 \\
$\log(gf_{40})$   &$-$1.458&$-$1.90  &$-$1.416  &$-$1.362    &$-$0.463  \\
$\log(gf_{48})$   &$-$1.416&$-$1.416 &$-$5.67   &$-$0.463    & $-$4.800 \\
$a'$              &        &         &0.949     &            & 0.5   \\
$d'$              &        &         &5.00      &            & 5.0   \\
$q$               &        &         &0.185     &            & 0.2    \\ \hline
\end{tabular}
\label{tab:10AB}
\end{table}

There is no question that the fit for $\lambda$8498 is better with
the model that assumes isotopic stratification, as claimed by RKB.
We shall make an overall assessment after the two stronger IRT
lines, and the Ca {\sc ii} K-line have been discussed.

\subsection{The $\lambda$8542 line}

Neither of the stronger IRT lines were examined in RKB's study.
The $\lambda$8542 line is the strongest, and was
generally unavailable on UVES spectra prior to November 2004.
The intrinsic strength of $\lambda$8542 is nearly 9 times
greater than that of $\lambda$8498.  One therefore expects to
see better-developed wings.
This should give an increased chance
of detecting a wavelength shift---between core and wings---if
the core is primarily due to the heavy isotope while
the wings are
formed deep, where the light isotope dominates.

\begin{figure}
\includegraphics[width=84mm,height=84mm,angle=0]{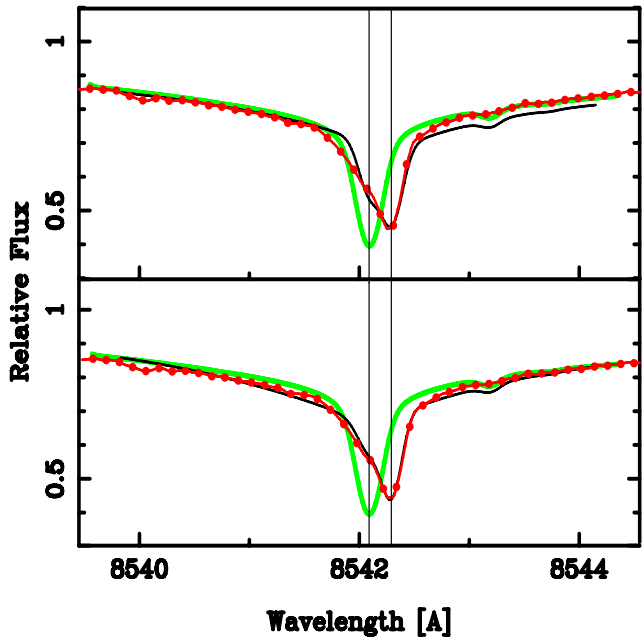}
 \caption{The $\lambda$8542 line in 10 Aql (UVES
4 August 2008).  The upper figure: elemental stratification;
lower figure: elemental {\it and} isotopic stratification.
Coding as in Fig.~\ref{fig:8498c}.
The thick gray line (green in online version) is for a
profile with pure $^{40}$Ca; contrast the behavior of the
wings for the profiles.
Relevant parameters for the fit are in Tab~\ref{tab:10AB}.
\label{fig:8542c}}
\end{figure}

The upper part of  Fig.~\ref{fig:8542c} shows an automatic (simplex)
fit (black) to the observed profile, assuming only elemental
stratification.
The constant $^{48}$Ca/$^{40}$Ca ratio is about 6 to 1.
Note the greater strength of the calculated red wing,
and compare the wing with the observed (dark gray with dots)
and profile for pure $^{40}$Ca (thick lighter gray).  This
behavior was noted by Ryabchikova and her coworkers
(e.g. RKB) as an indication that the wing was formed by
a normal (mostly $^{40}$Ca) mixture.

The lower calculation (black) assumes the $^{48}$Ca is in
a high cloud, with parameters ($g_{48}(x)$) given in
Tab.~\ref{tab:10AB}.  Trial and error improvements were
made after an automatic fit to obtain the profile shown.

If we compare the upper and lower parts of Fig.~\ref{fig:8542c},
we see the same general features as shown in
Fig.~\ref{fig:8498c} for the weaker line, $\lambda$8498.
Without isotopic stratification, one cannot get enough
absorption in the violet
wing without exceeding the observed minimum at
the centroid of the absorption for
$^{40}$Ca (left vertical lines in
Figs.~\ref{fig:8498c} and \ref{fig:8542c}.
The isotopically stratified model can accomplish
this because it reduces the amount of $^{40}$Ca in the
upper atmosphere, where the core of the line is formed.

The parameters for the two lines in Tab.~\ref{tab:10AB}
in columns 5 and 6 for the two lines differ somewhat
for the stratification ($g(x)$, and $g_{48}(x)$).
It difficult to judge how meaningful these differences
are.
Column densities may be more meaningful.
We discuss specific results in \S\ref{sec:NH}.

\subsection{The $\lambda$8662 line}

The $\lambda$8662 line (not shown) is intrinsically
some 55\% as strong
as $\lambda$8542.  It can be fit with parameters similar to
those in Tab.~\ref{tab:10AB} for the other two IRT lines.
The fit in the near, violet
wing is complicated by a strong Fe I line, at 8661.90~\AA.
If we adjust the iron abundance, or the appropriate $gf$-value
for that feature, a fit
using only elemental stratification
shows the same general features as the other lines of the
IRT.  Some absorption is missing in the violet wing, and
the red wing is too deep.

For the present, we conclude isotopic stratification
is the simplest way to explain the IRT profiles in 10 Aql.

\subsection{The Ca {\sc ii} K-line in 10 Aql;
overall column densities\label{sec:NH}}

\begin{figure}
\includegraphics[width=80mm,height=84mm,angle=00]{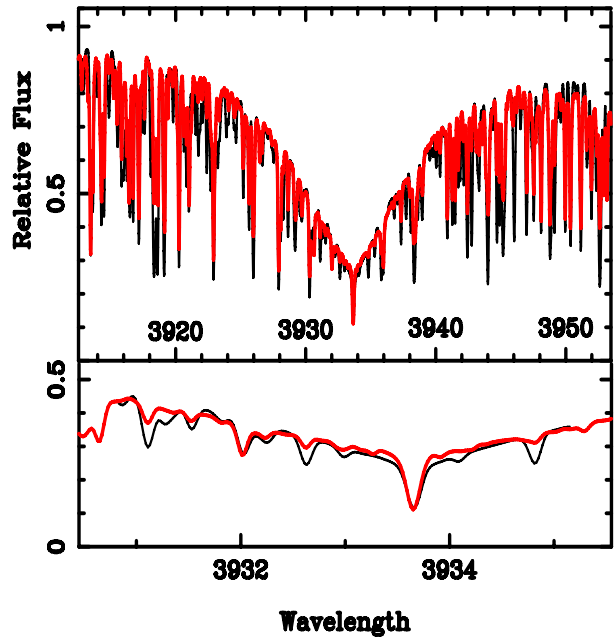}
\caption{Ca {\sc ii} K-line fits in 10 Aql. Observation:
gray (red online; UVES 8 October 2001), calculation: black.
Both plots are centered at
$3933.00$~\AA\,  The K-line core ($3933.66$~\AA) is therefore
noticeably displaced in the lower figure, which shows
the inner part of the fit.  The
stratification parameters are $a = 2.2$,
$b = 3.0\cdot 10^{-6}$, $d = 0.1$,
and ${\rm Ca}/N_{\rm tot} = 2.0\cdot 10^{-4}$.
\label{fig:10AqlK2}}
\end{figure}

Fig.~\ref{fig:10AqlK2} shows our fit to the Ca {\sc ii}
K-line in 10 Aql including a close up of the fit in the
core.  Relevant stratification parameters are given
in the caption.

Tab.~\ref{tab:cold} compares the column density of
the K-line with those for the IRT.

\begin{table}
 \centering
  \caption{Logarithmic column densities for
  Ca II lines in 10 Aql.}
  \begin{tabular}{l l l l l}  \hline
Isotope$\backslash\lambda$&8498&8542  &8662 &K-line\\ \hline
\multicolumn{5}{c}{Elemental strat. only}  \\ \hline

48            &18.93  &19.34 &18.85&     \\
40            &18.90  &18.44 &18.07&     \\
40+48         &19.22  &19.40 &18.92&18.93 \\ \hline
\multicolumn{5}{c}{Elemental + isotopic strat.} \\  \hline
48            &14.39  &14.46 &13.38&        \\
40            &19.17  &19.02 &19.29&  \\
40+48         &19.17   &19.02 &19.29&18.93   \\  \hline
\end{tabular}
\label{tab:cold}
\end{table}

We know of no previous work that has assembled column densities
for stratified atmospheres.  Thus, we have no basis for judging
how well the values for features all arising from Ca {\sc ii}
should agree with one another.  The totals for the
calculations with elemental stratification only differ by
a maximum of 0.48 dex, or a factor of about 3.
When isotopic
stratification is added, the maximum spread is only slightly
less, 0.36 dex or a factor of 2.3.

\section{HD 122970}
\label{sec:122970}

Handler and Paunzen (1999) discovered the roAp nature of
HD 122970.  It was among the objects studied for elemental
{\it and} isotopic stratification by Ryabchikova (2005).
In Paper I, we gave shifts for $\lambda\lambda$8498 and 8662
of 0.13 and 0.19~\AA, respectively.  New measurements of all three
IRT lines have been made based on spectra obtained in August
2008.  They yield the following shifts: 0.15, 0.18, and 0.19~\AA\,
for $\lambda\lambda$8498, 8542, and 8662 respectively.
Differences for the measurements in common are 0.02 and 0.00~\AA,
in good agreement for broad, asymmetrical lines.  The tendency
to measure the weaker line of the triplet at a smaller shift
than the stronger lines is repeated in the remeasurement.

\begin{figure}
\includegraphics[width=80mm,height=84mm,angle=0]{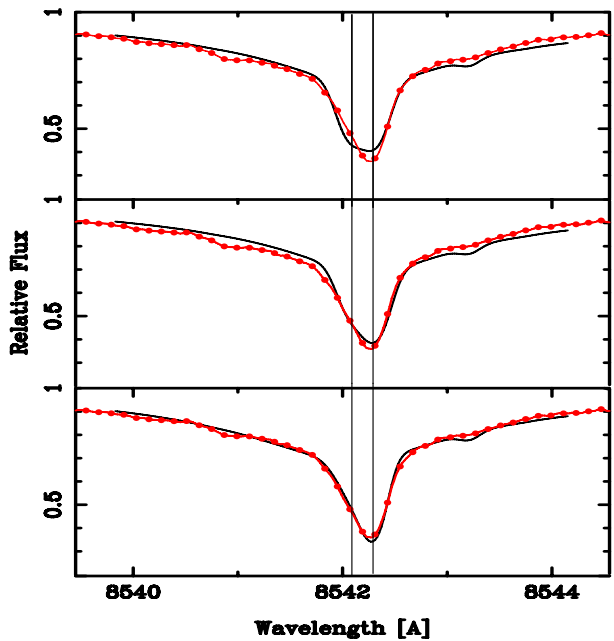}
 \caption{The strongest IRT line, $\lambda$8542 line in
HD 122970 without (top, and center)
and with (below) isotopic stratification
(UVES 3 August 2008).  Parameters for the fits are given
in Tab.~\ref{tab:970B}.  The upper fit was done using the
simplex code.  The center plot is a trial and error (t\&e)
fit, starting from the simplex parameters.  The lower
plot is an unmodified simplex solution (see text).
\label{fig:970B}}
\end{figure}

Fig.~\ref{fig:970B} shows
fits of theoretical spectra to the new observations, specially
reduced by FG.  Our model is based on the parameters of
Ryabchikova et al. (2000).

\begin{table}
 \centering
  \caption{Parameters  of $\lambda$8542
   fits for HD 122970 with uniform
  and isotopic stratification.  Cols. 2  and 3 refer
  to the top and center plots of Fig.~\ref{fig:970B}.
  Col. 4 gives parameters for the lower plot of that
  figure.}
\begin{tabular}{l r r r}    \hline
Param.            &uniform  & uniform & iso-strat \\  \hline
                  &simplex  &  t\&e    &simplex    \\
$a$               &6.361    &6.300    &7.342      \\
$\log(b)$         &$-$3.455 &-3.301   &$-$4.302     \\
$d$               &0.562    & 0.562   &0.752      \\
log(Ca/$N_{\rm tot})$&$-$4.947 &$-$4.824 &$-$5.213     \\
$\log(gf_{40})$   &$-$0.708 &$-$1.400 &$-$0.316     \\
$\log(gf_{48})$   &$-$0.460 &$-$0.460 &$-$0.460     \\
$a'$              &         &         &1.500       \\
$d'$              &         &         &6.488       \\
$q$               &         &         &0.009       \\ \hline
\end{tabular}
\label{tab:970B}
\end{table}

To make a judgement on whether isotopic stratification is
indicated, compare the center and lower plots
of Fig.~\ref{fig:970B}.  Clearly, the lower fit is better.
In the middle plot
we see the effect pointed out by RKB for the weaker $\lambda$8498
line.  The red
wing is below the observations, while the violet wing is above.
This is explained by the hypothesis of a constant isotopic ratio,
which makes too great a contribution to the red wing from
the $^{48}$Ca.

A better fit is shown in the lower part of the figure, where
the wings are primarily due to $^{40}$Ca.  The effect is not large,
but it is consistent with the effect shown in several papers
by Ryabchikova and coworkers, which dealt only with
the weaker $\lambda$8498 line.  This consistency argues against
the possibility that the improved fit is simply due to the
additional parameters of the isotopically stratified model.

With isotopic stratification, the simplex calculation puts
the maximum of the $g_{48}$ function {\it above} the top
layer of our model [$\log(\tau_{5000}) = -5.4$].  We discussed
a similar result
in \S\ref{sec:9810} for $\lambda$8498 in 10 Aql.  The
effect was already noted by Ryabchikova (2005).
We have noted the need for a study including a
hyperextended atmosphere (cf. \S\ref{sec:hypex}),
and non-LTE.
We find that an equally good fit to the observed
profile may be made if we modify the simplex parameters
slightly, as we discussed in \S\ref{sec:9810}.
In particular, we used $a' = 6.0$, $d' = 5.0$, and
$q = 0.009$.  We get an excellent fit, when we also
multiply the $g_{48}(x)$ by 0.01.

With the latter parameters, we find a column density
$\rm \log(^{48}CaH) = 14.38$, and
$\rm \log(^{40}CaH) = 18.26$.  This relatively very
low column density for $^{48}$Ca should be contrasted with
the value that follows from the parameters of the uniform
trial and error solution: 18.33.  Here, most of the calcium
is assumed to be in the heavy isotope, and the total column
density is essentially the same as for $^{40}$Ca with the
isotopically stratified model.

The logarithm of the total
column density
of massive particles is 23.97, so the overall
$\rm \log(Ca/N_{tot})$ value is $-$5.64, close to the
corresponding solar value, $-$5.65.

\section{Gamma Equ}

Frequent statements may be found in the literature of CP stars
that $\gamma$ Equ and 10 Aql have very similar spectra (cf.
Wolff 1983, Ryabchikova, et al. 2000).  Probably, the idea
goes back to a comment by Bidelman (remark to CRC), whose
careful intercomparisons of high resolution spectra of
CP stars in the 1960's were (and are)
both highly regarded and well
known to those who study the spectra of these stars.
Because of its close association with 10 Aql,
we include $\gamma$ Equ in the present study,
even though the average IRT shifts (0.13\,\AA) are not quite
large enough for it to be included in Tab.~\ref{tab:bigshifts}.
Subsequent
work has shown that neither the abundances nor the spectra
are identical, though the spectra are much more like one
another than to many other cool CP stars
(cf. $\beta$ CrB, HR 7575).

In Paper II, we fit the $\lambda$8542 line of the IRT.  That
calculation was made without the current refinements that
take Paschen confluence into account (cf.
Paper III, Appendix A).  The additional continuous opacity
from ``dissolved'' upper levels accounts for a
difference in line depth of the order of 0.05 of the
continuum, in the line wings.  Additionally, the effective
oscillator strength of P15 is reduced, because some of
the line opacity is now (quasi-) continuous opacity.
This could account for
the difference of a factor of 4 in the ${\rm Ca'}/N_{\rm tot}$
values shown in Tab.~\ref{tab:params}.  While no adjustment
to the 10 Aql continuum was made for the specific purpose of
fitting the IRT lines, we must admit that the uncertainty
in placement of the continuum is of the order of several
per cent.

\begin{figure}
\includegraphics[width=80mm,height=84mm,angle=0]{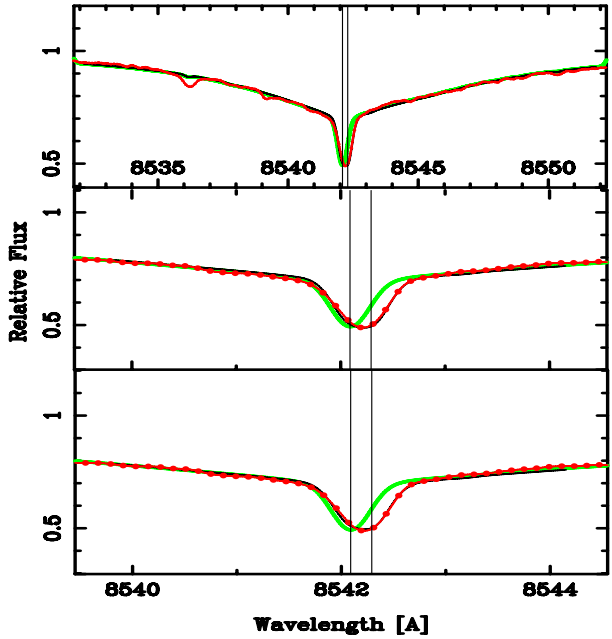}
 \caption{The $\lambda$8542 line in $\gamma$ Equ (UVES spectrum
from 18 September 2005).
Parameters for $g(x)$
are given in Tab.~\ref{tab:params}.  They apply to the upper
and center plots.  The lower plot is an automated fit assuming
isotopic stratification.
The solid black line is
the calculation.  The observed spectrum
is the thin line in dark gray (red in
online version).
The broad, light gray line (green in
online version) shows a fit assuming all of the calcium is
$^{40}$Ca.
\label{fig:geq_8542}}
\end{figure}

\begin{table}
 \centering
  \caption{Parameters of the $\lambda$8542 fit in $\gamma$Equ}
  \begin{tabular}{l c  c}
  \hline
$\lambda$8542 &Current       &Paper II\\
              &work              &Fig. 7 \\  \hline
Ca'/$N_{\rm tot}$&$1.2\cdot 10^{-4}$&$3.0\cdot 10^{-5}$ \\
$\log(gf)_{40}$&$-$0.46           &$-$0.46    \\
$\log(gf)_{48}$&$-$0.36           &not used \\
$a$           &6.7               &6.7 \\
$b$           &$1.5\cdot 10^{-5}$&$10^{-4}$ \\
$d$           &0.75              &1.0 \\  \hline
\end{tabular}
\label{tab:params}
\end{table}



\begin{figure}
\includegraphics[width=80mm,height=84mm,angle=0]{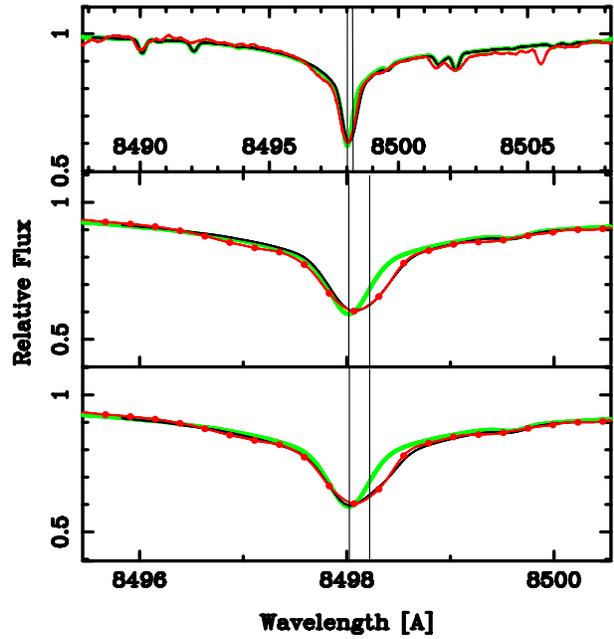}
 \caption{The $\lambda$8498 line in $\gamma$ Equ (UVES
spectrum from 18 September 2005).
Parameters for $g(x)$, which apply to the upper two
plots
are: ${\rm Ca'}/N_{\rm tot} = 7.0\cdot 10^{-4}$,
$a = 6.7$,
$b = 2.5\cdot 10^{-6}$, $d = 0.30$, $\log(gf)_{40} = -1.12$,
$\log(gf)_{48} = -1.32$.  The lower plot is an automated
fit assuming isotopic stratification.
Color coding as
in Fig.~\ref{fig:geq_8542}.
\label{fig:geq_8498}}
\end{figure}

Fits to the $\lambda$8498 line are shown in
Fig.~\ref{fig:geq_8498}.  The stratification parameters
and value of ${\rm Ca'}/N_{\rm tot}$ are somewhat different
from those used for the $\lambda$8542 line.  A calculation
using the same parameters fits reasonably well in the core
and far wings, but is much too strong in the near wings.
Optimum parameters are given in the figure.

The third line of the IRT, $\lambda$8662, is well fit by
the same stratification parameters as the stronger
$\lambda$8542 line, but with
${\rm Ca'}/N_{\rm tot} = 3.0\cdot 10^{-4}$, and
$\log(gf)_{40} = -1.00$, and $\log(gf)_{48} = -0.72$.
The differences may not be significant.  When we fit
the IRT lines in Paper III, we found the same stratification
fit the two stronger lines, while the weaker $\lambda$8498
line, required significantly different stratification parameters.

Note the good fits for the red wings in the upper two
plots for both Figs.~\ref{fig:geq_8542}
and ~\ref{fig:geq_8498}.
It does not appear necessary to invoke {\it isotopic}
stratification to
account for the IRT profiles in $\gamma$ Equ.

\subsection{The Ca {\sc ii} K-line in $\gamma$ Equ\label{sec:hypex}}

In Paper II, we noted (\S 7.4) that the same parameter set
that fit the $\lambda$8542 line also ``accounts quite well
for the Ca {\sc ii} K-line profile.''  This situation
must be reexamined because
of the current use of extra
opacity from dissolved Paschen lines.
We find that slightly modified parameters
(cf. Tab.~\ref{tab:params}, Col. 2) provide a good fit
to the K-line: ${\rm Ca}/N_{\rm tot} = 9.0\cdot 10^{-5}$,
and $d = 0.65$.  The $a$, and $b$ parameters are the same.

Ryabchikova, et al. (2002, RPK) examined the
Ca {\sc ii} K-line in $\gamma$ Equ in a study that employed
a hyperextended atmosphere, to $\log(\tau_{5000}) = -10$.
Since the center of the K-line saturates in the first depth
of our atmosphere [$\log(\tau_{5000}) = -5.4$], such an
extension would be appropriate.  We have experimented with
similar models, and find they give essentially
the same  profiles to
the one currently used, provided the temperatures are
appropriately adjusted at the shallowest depths.  Since
such atmospheres are poorly constrained by radiative
equilibrium in LTE, we use our standard model here.

\begin{figure}
\includegraphics[width=85mm,height=84mm,angle=0]{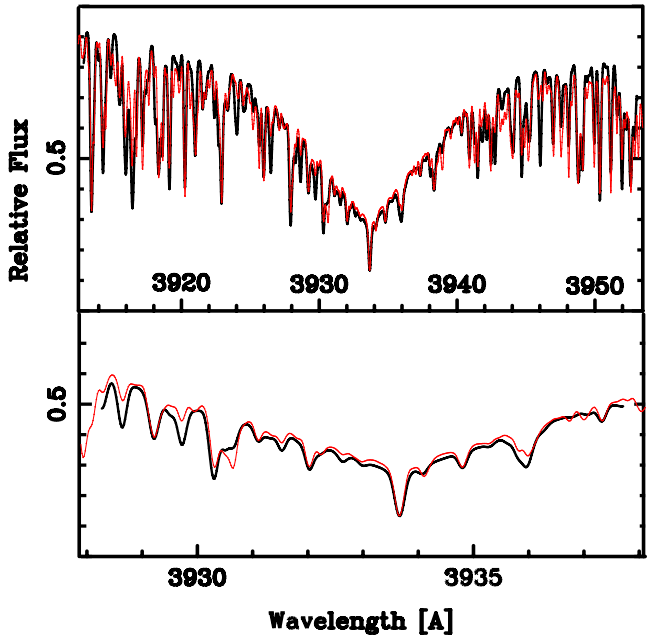}
 \caption{The Ca {\sc ii} K-line in  $\gamma$ Equ (UVES
 spectrum from 18 September 2005).
Parameters for $g(x)$
are: ${\rm Ca}/N_{\rm tot} = 9.0\cdot 10^{-5}$, $a = 6.7$,
$b = 1.5\cdot 10^{-5}$, $d = 0.65$, $\log(gf) = -0.100$.
The central part of the profile is displayed in the lower
part of the figure.
\label{fig:2Kgeq}}
\end{figure}

\begin{figure}
\includegraphics[width=55mm,height=84mm,angle=270]{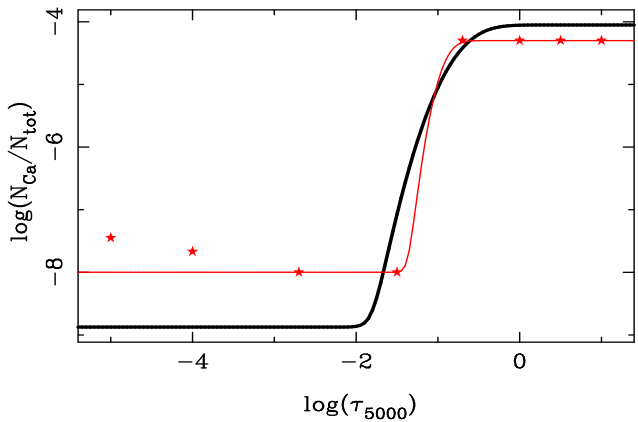}
 \caption{Stratification profiles leading to good fits
for Ca {\sc ii} K-line in $\gamma$ Equ (UVES 18 September 2005).
Parameters for the
thicker, black curve are given in the caption to
Fig.~\ref{fig:2Kgeq}.  The gray curve (red in online version)
is for the function $g(x)$, and used for a fit indistinguishable
from that shown in the lower part of Fig~\ref{fig:2Kgeq}.  The
parameters were
${\rm Ca}/N_{\rm tot} = 5.0\cdot 10^{-5}$, $\log(gf) = -0.100$,
$a = 30.0$, $b = 2.0\cdot 10^{-4}$, $d = 0.9$.
See text for the meaning of the filled stars.
\label{fig:sprofs}}
\end{figure}

Fig.~\ref{fig:2Kgeq} shows a fit to the K-line in
$\gamma$ Equ, with a closeup of the core.  The
parameters are indicated in the caption.
An equally good fit may results from
parameters, chosen to approximate those shown for calcium
in RPK's Fig. 3.  The two stratifications and relevant
parameters are  given in
Fig.~\ref{fig:sprofs}.  The filled stars indicate the
stratification profile used by RPK, which deviates at the
highest layers from the approximation.

\section{Stars with weaker calcium lines: Dominant $^{48}$Ca}

\subsection{HgMn stars}
\begin{figure}
\includegraphics[width=55mm,height=84mm,angle=270]{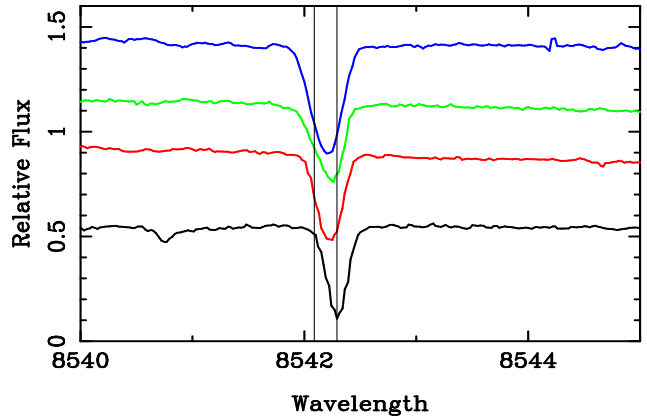}
 \caption{The strongest IRT line, $\lambda$8542 in 4 HgMn
stars.  Arbitrary vertical displacements of roughly normalized
spectra are made for display purposes.  The stars,
along with the dates of the UVES exposures are from top
down:
HD 29647 (5 August 2008),
HR 1800 (19 September 2005),
HR 7245 (18 October 2005), and
HR 7143 (20 September 2005).  The vertical
lines mark the positions of pure $^{40}$Ca and pure $^{48}$Ca.
\label{fig:4Mn}}
\end{figure}

Several CP stars with large isotopic shifts have rather
weak Ca {\sc ii} and/or IRT lines--certainly relative to HR 1217.
RKB do not discuss any of the HgMn
stars, which also show varying
isotope shifts.  Several examples are shown in Fig.~\ref{fig:4Mn}

Only
HR 7143 (HD 175640) shows the full shift corresponding to
$^{48}$Ca.  This star was examined for elemental
stratification  by Castelli and Hubrig (2004) and
Thiam et al. (2008), who reported some evidence for
stratification from metallic lines.  Elemental stratification
is generally accepted for emission lines common in the
red and infrared of HgMn and related stars (Sigut 2001).
The K-line of Ca {\sc ii} shows significant
wings, and might indicate stratification if it were
present.  However, Castelli's web site shows an excellent
fit with a non-stratified model:
http://wwwuser.oats.ts.astro.it.castelli\-/hd175640/
p3930-3936.gif

The slope of the  ``edge'' of the HR 1800 profile
is steeper on the red side than on the blue.  This shape is
common among the CP2 stars, as illustrated
elsewhere in this paper.  Presumably, the more shallow violet
slope is caused by an admixture of $^{40}$Ca.
Additional work is needed to investigate the question of
isotopic mixtures.  The profiles of HD 29647 and HR 7245
could arguably be primarily due to $^{46}$Ca, which has
a shift of 0.16~\AA\, relative to $^{40}$Ca.  Contributions
from lighter as well as heavier isotopes might be required.

One cannot rule out the possibility that
the $^{48}$Ca is in a high, stratified layer.
If this were the case, and the low photospheric
abundance of calcium were assumed very high,
the relative percentage of heavy calcium
above the photosphere could be much smaller
than it would appear from a naive examination
of the profile (see remarks below for the
IRT profiles in HR 5623).

Two other HgMn stars in Table A1 of Paper II
show large isotopic shifts: HR 6520 (HD 158704) and
HR 6759 (HD 165473).

\subsection{CP2 stars with weaker IRT lines}

Tab.~\ref{tab:bigshifts} contains CP2 stars
with large
isotope shifts (ca. 0.2~\AA) but
moderate or weak IRT lines: HR 5623 (HD 133792),
Przybylski's star (HD 101065), and HD 217522.
We discuss them in this order.
HR 5623 was the subject of the intensive
study of KTR, which introduced the vertical inversion
technique.
The latter two stars have minor absorption at best
that could be attributable to $^{40}$Ca.

\section{HR 5623 (HD 133792)}

\begin{figure}
\includegraphics[width=55mm,height=84mm,angle=270]{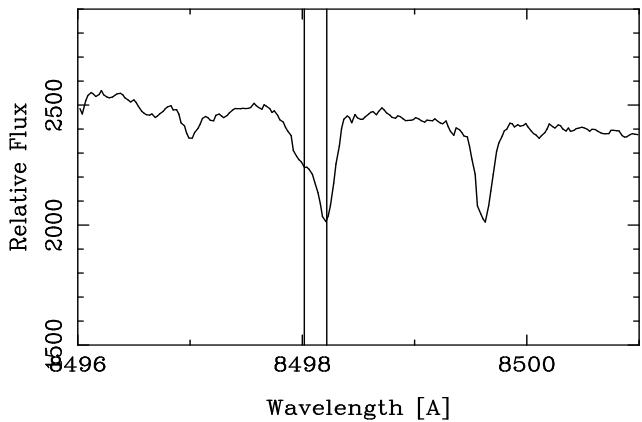}
 \caption{The $\lambda$8498 line in HR 5623 (UVES spectra
19 March 2006).  Vertical lines
mark the wavelengths for $^{40}$Ca and $^{48}$Ca.  While the
heavier isotope dominates the absorption, it does not
{\it necessarily} dominate the number of Ca ions above the
photosphere (\S\ref{sec:colden}, \S\ref{sec:colden23}).
\label{fig:A5623}}
\end{figure}

We devote special attention to HR 5623 because the
weakest IRT line, $\lambda$8498 has dominant absorption
at the position for $^{48}$Ca, with a significantly
smaller contribution from $^{40}$Ca.  This is shown
in Fig.~\ref{fig:A5623}.  It thus seemed possible that
the abundance profile for calcium in HR 5623 approximated
that in stars like
HR 7143 (HD 175640), where there is no obvious indication
of absorption from the lighter isotope {\it at all}.
Absorption from the stronger line of the $\lambda$8498
blend is roughly double that of the weaker component,
presumably due to $^{40}$Ca, so we might conclude the
relative numbers of the isotopes ``above the photosphere''
was roughly 2 to 1 in favor of $^{48}$Ca.
When absorption from the lighter isotope is entirely
missing, there is no way stratification can lead to
any conclusion other than $^{48}$Ca dominates.  It
might seem that is also the case when the $^{40}$Ca
contribution is relatively weak; stratification would
not significantly change the apparent dominance
of heavy calcium.  However, we shall see that this is
not necessarily the case (\S\ref{sec:colden23}).

The work by KTR and RKB on this star is based on UVES spectra
taken on 26 February 2002.  Additional spectra were obtained
on 27 January 2005, after reconfiguration of the instrument
which move the order gaps away from the  IRT
lines.  IRT shifts from Paper II were +0.18~\AA~\ for both
$\lambda\lambda$8498 and 8662.  The new measurements
yield shifts of +0.19, +0.20, and +0.20~\AA~\ for
$\lambda\lambda$8498, 8542, and 8662 respectively.

\subsection{Effective temperature and gravity}

KTR {\it fixed} $T_{\rm eff} = 9400$K, and
$\log(g) = 3.7$ prior to carrying out their vertical
inversion calculations.
The model parameters were
based on Str\"{o}mgren and H$\beta$ photometry,
and the Moon \& Dworetsky (1985) calibration
as implemented by Rogers (1995, TEMPLOGG).
Significantly, they adopted a reddening $E(B-V) = 0.09$,
which they state ``...follows from the reddening maps by
Lucke (1978) and high-resolution dust maps by Schlegel,
Finkbeiner \& Davis (1998)."
Additionally,
they fit H$\alpha$ and H$\beta$ profiles.


The assumed excess, $E(B-V) =  0.09$, follows
directly from a standard absorption and reddening law
(see Eq. 3.66 of Binney and Merrifield 1998),
and the Hipparcos parallax of 5.87 mas.
If the color
excess is correct it supports the
assumed temperature of 9400K.

There is reason to
suspect the effective temperature may be several hundred
degrees cooler.    A code
kindly provided to CRC by B. Smalley
(private communication), but based on the
Moon-Dworetsky (Moon 1984) work gives 8960 or 8900K, depending
on whether the reduction is done with Class 5 (A0-A3 III-V),
or Class 6 (A3-F0 III-V).  The reddenings $E(b-y)$, are
$-$0.001 and +0.032 respectively.

We may make an estimate of the reddening from the interstellar
sodium lines, with the help of the work of Munari and Zwitter
(1997).  The equivalent width of the Na D$_1$ is line difficult
to measure precisely, because it is partially blended with
the stellar line.  We estimate 96 m\AA.  From Munari and
Zwitter's Figure 1, one sees that this equivalent width
would correspond to $E(B-V)$ values ranging from 0.00 to
perhaps 0.10.  These authors also provide an empirical fit
to the relation between $E(B-V)$ and the equivalent widths
of Na D$_1$ as well as K I $\lambda$7699.  We fit a quadratic
to the first 5 values of their Table 2, and obtain $E(B-V) = 0.026$
from the Na line.  The K I feature is arguably present.  We
estimate an equivalent width of 6.7 m\AA\,, which would
correspond to $E(B-V) = 0.020$.

Paunzen, Schnell, and Maitzen (2006) give the excess in
the Geneva system as
$E(B2-V1) = 0.63\cdot E(B-V)$.  With this reddening, the
Geneva colors (www.unige.ch/sciences/astro/an)
give $T_{\rm eff} = 8952$K and $\log(g) = 3.32$,
according to
a code kindly supplied by
P. North (cf. Kunzli, et al. 1997).
This assumes a metallicity ([Fe/H])
of +1, and a reduction grid chosen automatically by
the code.

A spectroscopic determination of $T_{\rm eff}$ and
log($g$) may be made from the
equilibrium of Fe {\sc i} and {\sc ii} following the
method of Paper III, but using 4 temperatures,
8400, 8900, 9400, and 9900K, and 3
gravities, $\log(g) = 3.2$, 3.7, and 4.2.  The models
assumed abundances taken from KTR when available, otherwise
using solar values.  A reasonable compromise for the
microturbulence is 1 kms$^{-1}$.  The numerous slopes of
$\log({\rm Fe}/N_{\rm tot})$ vs. $\log(W_\lambda)$
are then sometimes slightly positive,
sometimes slightly negative.

Calculations were made using
an unstratified model and one stratified model with
parameters approximating the profile for iron of KTR,
but using a larger jump: $a = 20, b = 1.0\cdot 10^{-4},
d = 0.95$.  The larger jump was used because KTR found
a less than 1 dex jump.  We wanted to see the effect
of a stronger stratification.

The calculations provide combinations of $T_{\rm eff}$
and $\log(g)$ for which the Fe {\sc i} and {\sc ii}
give the same abundances.  Temperatures with equal abundances
from the two stages of ionization are given in Tab~\ref{tab:feeq}
for three surface gravities.
\begin{table}
 \centering
  \caption{$T_{\rm eff}$--$\log(g)$ pairs with
  Fe {\sc i} and Fe {\sc ii} in equilibrium.  Calculations
  with and without stratification (``strat,'' see text).
  Corresponding $\rm \log(Fe/N_{tot})$ are given in columns
  3 and 5.}
  \begin{tabular}{r r r r r}    \hline
$\log(g)$  & No Strat  &  Abund  &  Strat  &  Abund \\  \hline
3.2        &8914       & -3.67   & 9143    &  -2.59  \\
3.7        &9490       & -3.41   & 9653    &  -2.34 \\
4.2        &9834       & -3.21   & 9916    &  -2.30 \\ \hline
\end{tabular}
\label{tab:feeq}
\end{table}

The Geneva photometry and iron equilibrium agree on a low
temperature and surface gravity, and no iron stratification.
At least some implementations
of Str\"{o}mgren photometry concur.  A value of $\log(g)$ as
low as 3.2 would be unusual (cf. Hubrig, North,
Sch\"{o}ller \& Mathys 2007).

\begin{figure}
\includegraphics[width=55mm,height=84mm,angle=270]{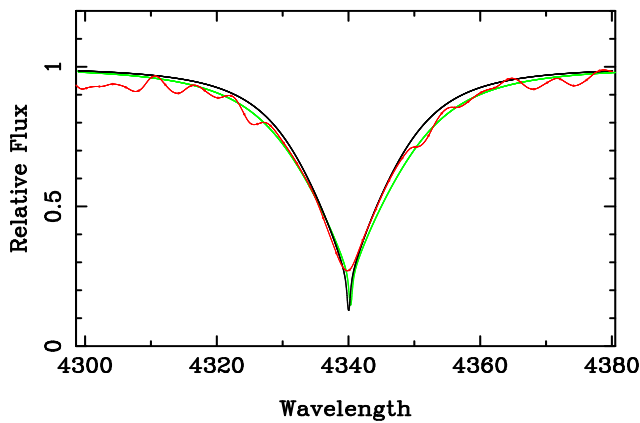}
 \caption{The FORS1 spectrum of H$\gamma$ (dark gray
with points) in HR 5623.  The light gray plot was made assuming
$T_{\rm eff} = 9400$K, and $\log(g) = 3.7$.  The black
curve was made using
$T_{\rm eff} = 8900$K, and $\log(g) = 3.2$.
The fits are comparable in quality.
\label{fig:H5}}
\end{figure}

Unfortunately, the Balmer lines do not clarify the matter.
While KTR support their choice of temperature and gravity
by examining H$\alpha$ and H$\beta$, we find these profiles
are fit comparably well with $T_{\rm eff} = 8900$K, and
$\log(g) = 3.2$.  An example is illustrated in
Fig.~\ref{fig:H5}, based on theoretical profiles of
Stehl\'{e} \& Hutcheon (1999).
The stellar observations are of low resolution from FORS1
(Appenzeller 1998) which have some advantage
over the UVES for broad features.  However, we have also
examined rectified H$\alpha$ and H$\gamma$ profiles from
UVES spectra, and reach similar conclusions.

\subsection{Paschen lines}

In principle, the Paschen lines might also resolve the
ambiguity in temperature and surface gravity.  The situation
is hardly better than with the low Balmer lines.  Of the
three Paschen lines near the IRT, two are significantly
influenced by the series convergence.  In our calculations
P13 ($\lambda$8662) is not strongly affected.  We also made
calculations for P11 ($\lambda$8862) and P12 ($\lambda$8750).
All Paschen profiles used Lemke's (1997) tables, because
the newer Stehl\'{e} and Hutcheon calculations do not
go to large enough values of $N_e$ for high series members.
While series convergence is not a problem, photometry
and normalization probably is.
Calculated profiles for neither of the
favored models, gave good fits.  We suggest much of the
difficulty may lie with the observations, and note that
the 2006 UVES spectrum was not re-reduced by FG.

\begin{figure}
\includegraphics[width=55mm,height=84mm,angle=270]{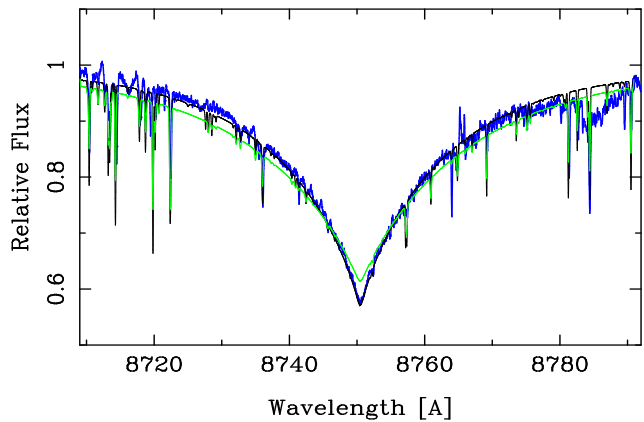}
 \caption{The HR 5623 UVES spectrum (19 March 2006)
of P12 (dark gray
with points, blue in online version).  The light gray
curve (green online) was calculated assuming
$T_{\rm eff} = 9400$K, and $\log(g) = 3.7$.  The black
curve was calculated using
$T_{\rm eff} = 8900$K, and $\log(g) = 3.2$.
The overall fit is better for the lower temperature,
for both core and wings.  No attempt has been made to
reconcile the atomic line spectrum with the observations.
\label{fig:P12}}
\end{figure}

Fig.~\ref{fig:P12} shows fits of calculated P12 lines
to the raw 2006 spectrum, counts vs. wavelength modified
only by: shifting the wavelength scale for a radial velocity
of 9.3  kms$^{-1}$, and dividing the counts by 2746.  The fit
certainly appears to favor the lower-temperature model, but
better observational material is needed.  The ``raw''
P12 profile differed only subtly from the normalized and
Fourier filtered version that we typically use.  However
the small differences were enough that none of the calculated
P12 profiles gave a good fit.  It seems that our (CRC)
best efforts at normalization actually degraded the
profile of this and perhaps other broad features.

\subsection{Calcium}

\begin{figure}
\includegraphics[width=85mm,height=84mm,angle=0]{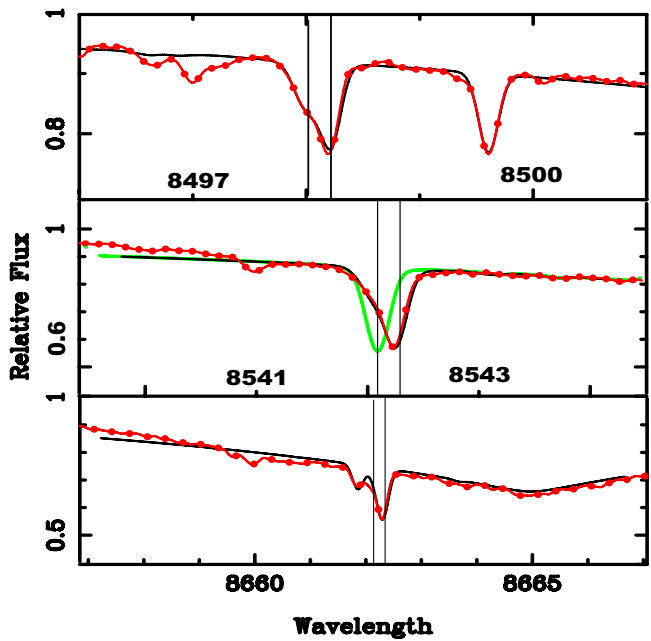}
 \caption{The IRT in HR 5623 (UVES spectra 19 March 2006).
 The calculated profiles (solid black); observations are
 dark gray (red online) with dots.  The light gray profile
 for $\lambda$8542 is calculated with $^{40}$Ca only.
 There is no indication, as in Fig.~\ref{fig:wing9862}
 that pure (or dominant) $^{40}$Ca would give a better fit
 in the red wing.
 Adjustments (see text) have been made to the Paschen
 slopes to fit the appropriate Paschen calculations
 (see text).
\label{fig:5623_3}}
\end{figure}

We see little basis in the IRT
{\it profiles alone} for assuming
that calcium is isotopically
stratified in the atmosphere of
HR 5623.
This  is clear from RKB's Fig. 8 (uppermost plot),
as well as Fig.~\ref{fig:5623_3}.  Indeed, the
evidence for any stratification of calcium at all
is not strong.  This is shown in the lower plot of
our figure, as well as Tab.~\ref{tab:Ca1Ca2}.
Reasonable fits to all three IRT profiles may be
obtained with any one of three contending stratification
assumptions: none, elemental, and isotopic.

To show the relative insensitivity of these weak lines
to model assumptions, we show the $\lambda$8498 fit to
an isotopically stratified model, the $\lambda$8542 fit
with elemental, but not isotopic stratification, and
the $\lambda$8662 fit with no stratification.  Fits of
all three lines with any of the three model assumptions
closely resemble those shown in the figure.

None of the IRT lines show an
indication of displaced wings that would indicate a
$^{40}$Ca-dominated deep photosphere (cf.
Fig.~\ref{fig:wing9862}).
Small
vertical adjustments of the observed Paschen wings were necessary
to achieve the fits shown for the upper plots.  We have
already noted photometric uncertainties in this region.


The difficulty in extracting a stratification profile
for calcium is well illustrated in KTR's Figure 9.  Of
the seven lines used, four are only marginally above
the level of the noise.  Useful information is probably
contained in the Ca K-line, but there are difficulties
with Ca {\sc i} $\lambda$4227, and Ca {\sc ii} $\lambda$3159.
The former is badly blended with Cr I, while much of
the discrepancy illustrated for the latter may be resolved by
taking a cooler model.  Additionally, there are continuum
problems in the region of this line.

RKB used a different, but partially overlapping,
set of Ca {\sc i} and {\sc ii} lines
from KTR, and obtained somewhat different stratification
parameters.  We have approximated them with our $g(x)$
function with parameters given in Tab.~\ref{tab:ktrrkb}.

\begin{table}
\centering
\caption{Abundances and $g(x)$ parameters for
approximations to KTR and RKB stratification profiles.}
\begin{tabular}{l  l  l}    \hline
Parameter       & KTR               &  RKB             \\  \hline
$a$             &6.0                & 30                  \\
$b$             &$1.29\cdot 10^{-4}$&$3.16\cdot 10^{-3}$  \\
$d$             &0.0                & 0.6                  \\ \hline
\end{tabular}
\label{tab:ktrrkb}
\end{table}

Only the Ca {\sc ii} K-line profile approaches the strength
needed to see stratification from the line shape alone.
Even for the K-line, the case for stratification is marginal.
Attempts to derive the stratification from the profile should
also consider contributions from an interstellar component.

KTR claimed a Ca overabundance of 1.4 dex in the deep
photosphere (see their \S 6).  Assuming a solar
$\rm \log(Ca/N_{\rm tot} = -5.73$, KTR's assumption for
this value would be $-$4.32.  This should be compared to
RKB's value (see their Tab. 4)
$\rm \log(Ca/N_{\rm tot})_{\rm lo} = -5.6$, which is in
better agreement with our values (cf. Tab.~\ref{tab:Ca1Ca2}),
depending on the assumed model.


Little information is available from Ca {\sc i}.
Most of the lines
are very weak or badly blended.  Only the resonance line,
$\lambda$4227 is of modest strength (ca. 32 m\AA).  We calculate
abundances for this line, assuming that it is blended with
Cr {\sc i} $\lambda$4226.75, and that the chromium abundance is
fixed at
$\rm Cr/N_{tot} = 1.76\cdot 10^{-4}$.  Similarly, we adopted
a measured 225\,m\AA\, for the K-line, and computed abundances from
it, including blends with Cr and Fe I, which made only small
differences in the resulting abundances.  Results for
three models, with and without the RKB stratification are
given in Tab.\ref{tab:Ca1Ca2}.

\begin{table}
 \centering
  \caption{Logarithmic abundances
  (log(Ca/$N_{\rm tot}$) from 32 m\AA\, $\lambda$4227,
  and 225 m\AA\, K-line. For each model, the upper abundance
  id for Ca {\sc i} and the lower for Ca {\sc ii}
  Columns marked ``diff'' are (Ca {\sc i} $-$ Ca {\sc ii}).
  The last three rows give abundances that yield fits to
  the IRT lines, but only for the 8900K, $\log(g) = 3.2$ models.
  In the case of the stratified model, the
  abundances refer to the deep photosphere.}
  \begin{tabular}{l  l  l r  l r}    \hline
$T_{\rm eff}$(K)& $\log(g)$ &strat&diff&no strat&diff \\  \hline
 9400           &3.7  &$-5.41$&     & $-6.49$&    \\
                &     &$-6.02$& 0.61& $-7.19$& 0.70\\
 8900           &3.7  &$-6.46$&     & $-7.74$&  \\
                &     &$-6.52$& 0.06& $-7.50$&$-$0.24 \\
 8900           &3.2  &$-5.92$&     & $-7.31$&  \\
                &     &$-6.32$& 0.40& $-7.58$&0.27  \\   \hline
\multicolumn{2}{c}{$\lambda$(T8900/logg=3.2)} &strat&  &no strat \\ \hline
\multicolumn{2}{c}{8498}&$-$5.80 & &$-$8.21 \\
\multicolumn{2}{c}{8542}&$-$5.54 & &$-$7.94 \\
\multicolumn{2}{c}{8662}&$-$5.30 & &$-$7.82 \\  \hline
\end{tabular}
\label{tab:Ca1Ca2}
\end{table}

The best agreement between the K-line and Ca I $\lambda$4227
is for a stratified model with the
temperature-log($g$) pair (8900K--3.7).  Agreement is poorest
at the high temperature.  Deep photospheric abundances
agree reasonably well among the IRT lines; less well with
the K-line and $\lambda$4227.

\subsection{Column densities in HR 5623\label{sec:colden23}}

We calculated column densities for fits to the $\lambda$8498
profile, shown in Figs.~\ref{fig:A5623} and ~\ref{fig:5623_3}.
We get a good fit (not shown) using elemental (but not isotopic)
stratification, with parameters the same as in column 3
of Tab.~\ref{tab:ktrrkb}, but with ${\rm Ca'}/N_{\rm tot}$
smaller by a factor of 0.6.  This leads to a
logarithmic column density
for $^{48}$Ca of 16.51 (cgs)
and 16.00 for
$^{40}$Ca.

An equally good fit (Fig.~\ref{fig:5623_3}) was made
with both elemental {\it and} isotopic stratification.
In
this case, the assumed parameters for $g(x)$ were similar,
though not identical to those for elemental stratification
alone: $a = 30$, $d = 0.8$, $b = 3.16\cdot 10^{-10}$.  The
${\rm Ca'}/N_{\rm tot}$ value deep in the photosphere was set to
$5\cdot 10^{-6}$---this was essentially all $^{40}$Ca.  This
$g(x)$ essentially set all of the light isotope to zero
abundance for layers higher than $x = -1.5$.  The parameters
of $g_{48}(x)$ were $a' = 6$, $d' = 4$, and $q = 3$.  These
set the abundance of $^{48}$Ca to zero below $x = 0.5$.  By
$x = -1$, the $^{48}{\rm Ca}/N_{\rm tot}$ had reached its
maximum value of $4.12\cdot 10^{-9}$.

With isotopic stratification, even though the $^{48}$Ca feature
is stronger, the {\it column density} is much lower.  We
find a logarithmic column density for $^{48}$Ca
of 14.77, while that for
$^{40}$Ca is 17.22, a difference of 2.45 dex.

By column density, the relative fraction
of $^{48}$Ca is smaller, though of the same magnitude, as
the terrestrial fraction.  This surprising result was
noted in \S\ref{sec:122970}.  It happens
because the core regions of the line saturate very high
in the atmosphere, and its significance was pointed out
by RKB.
As far as the core is concerned,
the atmosphere below this point is invisible, and to some
extent, irrelevant.

The two possible structures examined here surely require
very different theoretical scenarios.

\section{Przybylski's Star}

The IRT lines in Przybylski's star appear to have the full shift
that would correspond to $^{48}$Ca.  Wavelengths from Paper II,
and new measurements from UVES spectra obtained on 1 Jan. 2006
are shown in Tab.~\ref{tab:przirt}.  It surely seems that
$^{48}$Ca dominates.

\begin{table}
 \centering
  \caption{Independent measurements of the wavelengths of
  IRT lines in HD 101065. Entries are shifts [\AA] from the
  assumed terrestrial wavelengths at
  8498.02, 8542.09, and 8662.14 \AA}
  \begin{tabular}{l  l  l l}    \hline
 Spectrum       & 8498      & 8542   & 8662    \\  \hline
 UVES/2002      &0.19       &        & 0.19      \\
 FEROS/2000     &0.16       &        & 0.18   \\
 UVES/2006      &0.21       & 0.18   & 0.20    \\  \hline
\end{tabular}
\label{tab:przirt}
\end{table}

\begin{figure}
\includegraphics[width=54mm,height=84mm,angle=270]{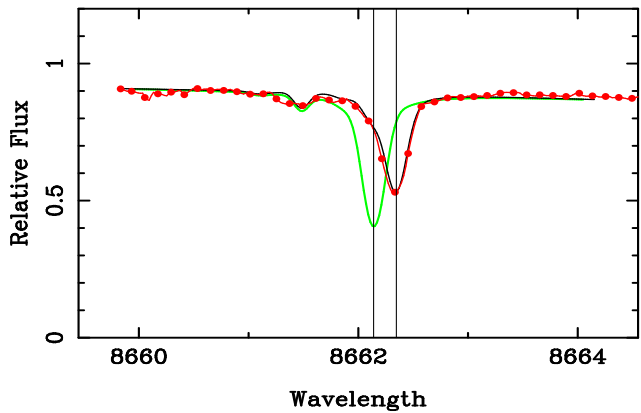}
 \caption{The $\lambda$8662 line in HD 101065
(UVES spectrum 14 January 2006).
Vertical adjustments of 3\% have been made
to achieve a fit to the wings of P13.  Calculations are in
black, observations dark gray with dots (red online).
Stratification
parameters are: $a = 6.7$, $b = 4.0\cdot 10^{-4}$, $d = 0.0$.
We used log(Ca'/$N_{\rm tot}) = -6.20$, and $\xi_t = 2$
kms$^{-1}$.  The ratio of $^{48}$Ca to $^{40}$Ca is 31.6.
The light gray curve was calculated with only $^{40}$Ca.
\label{fig:prz62}}
\end{figure}

We have not succeeded in finding a stratification profile
that will reconcile the deep photospheric abundances from
the IRT, the Ca {\sc ii} K-line, and Ca {\sc i} lines.
Nevertheless, the agreement among these features is
somewhat better with a provisional stratification than
without one.  The parameters are given in Fig.~\ref{fig:prz62}
for $\lambda$8662.

We now address the question of whether the heavy calcium
dominates the photosphere, or if it occurs in a high cloud
above a photosphere with primarily $^{40}$Ca.  We
apply the same test as used for HR 1217 and HR 5623, and
see if the wings of the IRT lines are better fit with
a shifted or unshifted theoretical profiles.  We do this for
the $\lambda$8662 line (Fig.~\ref{fig:prz62}).  There is no
indication that a pure $^{40}$Ca composition would yield
a better fit to the wings.  Compare this situation with that
illustrated for HR 1217 (Fig.~\ref{fig:wing9862}).  The
failure of pure $^{40}$Ca to account for the wings of
the strongest IRT line, $\lambda$8542, is similar, but the
photometry in the P15 wings is not good.

\begin{figure}
\includegraphics[width=54mm,height=84mm,angle=270]{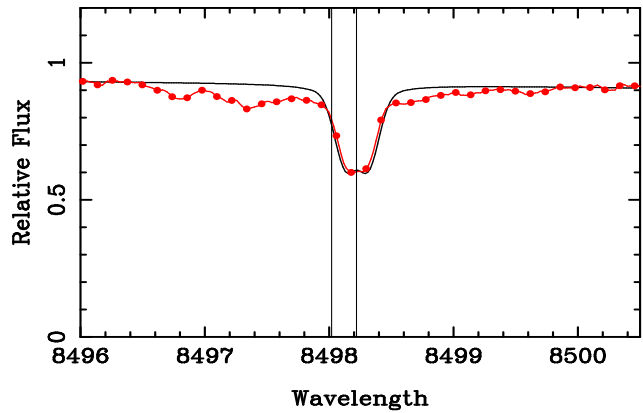}
 \caption{The $\lambda$8498 line in HD 101065 (UVES spectrum
14 January 2006).  Vertical adjustments of -2\% have been made
to achieve a fit to the wings of P16.  Calculations are in
black, observations dark gray with dots (red online).
Stratification parameters are as in Fig.~\ref{fig:prz62}.
Pure $^{48}$Ca was assumed.  See text for further discussion.
\label{fig:prz98}}
\end{figure}

The weakest of the IRT lines shows a slight Zeeman splitting.
It is calculated in Fig.~\ref{fig:prz98} assuming pure
$^{48}$Ca, and a transverse field of 2.9 kG.  The Zeeman
code was described in Paper III.
Paschen convergence was not included in this
calculation, but the effect is very small, because of the
low temperature of HD 101065.  The observed profile was
lowered by 2\% to fit the far P16 wings.

The cause of the broad local minimum on either side of the
$\lambda$8498 line is not known.  It is unlikely to be due
to wings, since the stronger $\lambda$8662
(Fig.~\ref{fig:prz62}) and $\lambda$8542 (not shown) do
not have extensive wings.    The abundance needed for the
fit shown was nearly 0.7 dex larger than needed to fit the
two stronger IRT lines.  It is unclear how meaningful this
is, since the function $g(x)$ changes from $5.2\cdot 10^{-4}$
to 0.500 between $\log(\tau_{5000}) = -1.0$ and 0.0.

We see little basis in
Fig.~\ref{fig:prz98} for assuming any contribution
of $^{40}$Ca to the profile.

\section{HD 217522}

HD 217522 is a roAp star discussed by Hubrig et al. (2002)
as a possible ``twin'' of Przybylski's star.  Gelbmann (1998)
showed that the star is both iron deficient, and at the
cool end of the CP star sequence.  He argued that this is
a general trend among roAp stars.

Measurements on the cores of the IRT lines in HD 217522 show
the full 0.20 \AA\, shift of $^{48}$Ca.  The lines themselves
are stronger than in Przybylski's star but do not show well
developed wings.  Paper II reported shifts of +0.18 and
+0.21~\AA, for $\lambda\lambda$8498 and 8662.  New measurements
of the spectrum obtained on 4 August 2008 yield shifts
of +0.20, +0.21, and +0.22~\AA, for
$\lambda\lambda$8498, 8542, and 8662, respectively.

\begin{figure}
\includegraphics[width=54mm,height=84mm,angle=270]{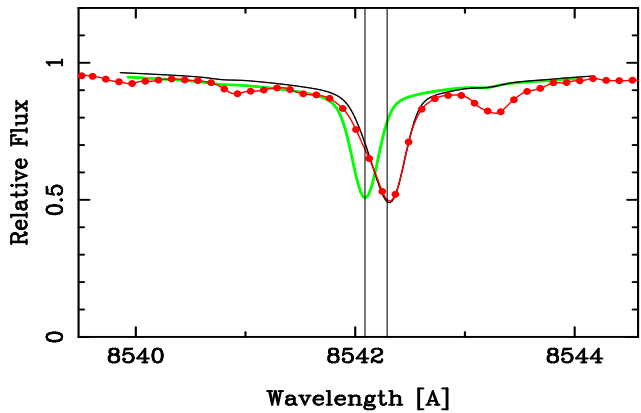}
 \caption{The $\lambda$8542 line in HD 217522
(UVES spectrum 4 August 2008).
Stratification parameters ($a = 20.0$,
$b = 1.0\cdot 10^{-4}$, $d = 0.1$) and the calcium abundance
were chosen to match
those used by RKB,
${\rm Ca'}/N_{\rm tot} = 1.58\cdot 10^{-5}$.  The abundance
applies to $^{48}$Ca.  A constant abundance ratio of
34.7 was assumed for $^{48}$Ca/$^{40}$Ca.  The light gray
curve (green online) was made with pure $^{40}$Ca.
\label{fig:B217}}
\end{figure}

The IRT can be fit equally well--arguably--with either an
isotopic stratification, or a stratification model with a
constant $^{48}$Ca/$^{40}$Ca ratio that is 10-20 to one.
We prefer the latter because fewer adjustable parameters
are required.
Fig.~\ref{fig:B217} shows a fit of the strongest line,
$\lambda$8542.  A calculation with pure $^{40}$Ca is shown
in light gray (green online).  We obtain quite similar
fits for the $\lambda\lambda$8498 and 8662 lines, with
similar though not identical stratification and abundance
parameters.
The plot resembles Fig.~\ref{fig:prz62}, where there is no
indication that the wings of the stellar feature would be
fit better with pure $^{40}$Ca.
Compare Figs.~\ref{fig:prz98} and ~\ref{fig:B217}
with Fig.~\ref{fig:wing9862}.

The same stratification profile and abundance fits the
Ca {\sc ii} K-line reasonably well.

\section{Discussion}

The profiles of strong lines in CP stars cannot be fit by
classical models that assume uniform element to hydrogen ratios
throughout the photosphere.  We support the assumption of
stratification to reconcile the discrepancies with observation,
employed by other workers.  These attempts are
not yet entirely satisfactory because somewhat different
stratification models are sometimes needed for lines of the
same element or ion.

Ideally, one may determine the
stratification parameters directly by comparison with the
observations.  Thus far workers have chosen a model,
and then
attributed deviations of observations from calculations
to stratification.
Possible errors in the
assumed model should be folded in to this procedure.
Such errors clearly influence the relative strengths
of neutrals and ions, as well as lines of a given species
with different strengths and excitation.  We discussed
one relevant example, HR 5623 (HD 133792), where we believe the
temperature was significantly overestimated.

We investigated RKB's bold hypothesis that fractionation of the
calcium isotopes could be observed in CP2 stars.
We support this claim for HR 1217, using different observational
material from their paper.
We confirm that good cases for such fractionation
can be made for 10 Aql and HD 122970.

One might argue that optimum fits with isotopic
stratification are only marginally better than the
optimum ones without it.  Certainly the additional
parameters and model flexibility of the isotopic
stratification would be expected to give an
improved fit.
However, the fits without
isotopic stratification show a consistent pattern
in a number of stars, including most studied by
RKB.  The calculated red wings are too strong,
indicating too much $^{48}$Ca in the deeper
atmosphere.


RKB wrote (\S\,6):
\begin{quote}
A simple interpretation of the anomaly observed in the Ca {\sc ii}
8498~\AA\, line core is to suggest that the heavy isotopes are
strongly enhanced and even dominant throughout the
atmospheres of some magnetic Ap stars.  However, our magnetic
spectrum synthesis calculations demonstrate that this hypothesis
is incorrect.
\end{quote}

The plots in
Papers I and II show that the IRT shifts vary from
small amounts to nearly the full amounts for $^{48}$Ca.  Only
a few stars show the full shifts.  Of these, HR 7143,
an HgMn star (Castelli and Hubrig 2004),
and two roAp stars, HD 101065
and HD 217522 show little or no
indication of the lighter isotope.  In HR 5623, one may
construct a model where $^{48}$Ca is not dominant, though
the most straightforward interpretation of the observations
is that it is.  In some HgMn stars, the symmetrical
profiles may suggest domination by isotopes of intermediate
mass.

The overall picture of isotope variations is complex.

\section{Acknowledgements}

We thank Drs. P. North and B. Smalley for computer codes,
and J. R. Fuhr of NIST for advice on oscillator strengths.
This research has made use of the SIMBAD database, operated
at CDS, Strasbourg, France.  We gratefully acknowledge
the use of ESO archival data, including the UVESPOP
data base.  Our calculations have made extensive use of
the VALD atomic data base (Kupka, et al. 1999).  We
appreciate the help of L. Sbordone and P. Bonifacio with
implementation of their version of Atlas 9.  Thanks
are also due to M. Netopil for help during observations
in August of 2008.

\section*{REFERENCES}

\hangpar Appenzeller, I., Fricke, K., F\"{u}rtig, W.,
et al. 1998, ESO Mess., 94, 1

\hangpar Babcock, H. W. 1958, ApJ, 128, 228

\hangpar Babel, J. 1992, A\&A, 258, 449

\hangpar Babel, J. 1994, A\&A, 283, 189

\hangpar Bagnulo, S., Jehin, E., Ledoux, C., et al. 2003,
ESO Mess., 114, 10

\hangpar Binney,J., Merrifield, M. 1998, Galactic Astronomy
(Princeton, N. J.: University Press)

\hangpar Bohlender, D. 2005, in Element Stratification in Stars,
40 Years of Atomic Diffusion, ed. G. Alecian, O. Richard,
S. Vauclair, EAS Pub. Ser. 17, 83

\hangpar Brage, T., Fischer, C. F., Vaeck, N., Godefroid, M.,
Hibbert, A. 1993, Phys. Scr., 48, 533

\hangpar Castelli, F., Hubrig, S. 2004, A\&A, 425, 263

\hangpar Cowley, C. R., Hubrig, S. 2005, A\&A, 432, L21 (Paper I)

\hangpar Cowley, C. R., Hubrig, S. 2008, MNRAS, 384, 1588
(Paper III)

\hangpar Cowley, C. R., Hubrig, S., Castelli, F. 2008, Contr.
Astron. Obs. Sk. Pl., 38, 291

\hangpar Cowley, C. R., Hubrig, S., Kamp, I. 2006, ApJS, 163, 393.

\hangpar Cowley, C. R., Hubrig, S., Castelli, F., Gonz\'{a}lez, F.,
Wolff, B. 2007, MNRAS, 377, 1579 (Paper II)

\hangpar Dworetsky, M. M. 2004, in IAU Symp. 224, The A-Star
Puzzle, ed. J. Zverko, J. \v{Z}i\v{z}novsk\'{y}, S. J. Adelman,
\& W. W. Weiss (Cambridge: Cambridge Univ. Press), p. 499.

\hangpar Gelbmann, M. J., Cont. Astron. Obs. Ska. Pl., 27, 280

\hangpar Handler, G., Paunzen, E. 1999, A\&AS, 135, 57

\hangpar Hubrig, S., Cowley, C. R., Bagnulo, S., Mathys, G.,
Ritter, A., Wahlgren, G. M. 2002, in {\it Exotic Stars as
Challenges to Evolution}, ASP Conf. Ser., 279, ed. C. A.
Tout \& W. Van Hamme, p. 365

\hangpar Hubrig, S., North, P., Sch\"{o}ller, M.,
Mathys, G. 2007, AN, 328, 475

\hangpar IMSL (R) Fortran 90 MP Library 3.0, \copyright
visual Numerics, Inc., 1998

\hangpar Kochukhov, O. 2007, in {\it Physics of Magnetic
Stars}, eds. I. I. Romanyuk \& D. O. Kudryavtsev, 109

\hangpar Kochukhov, O., Tsymbal, V., Ryabchikova, T.,
Makaganyk, V., Bagnulo, S. 2006, A\&A, 460, 831 (KTR)

\hangpar Kunzli, M., North, P., Kurucz, R. L., Nicolet, B. 1997,
A\&AS, 122, 51

\hangpar Kupka, F., Piskunov, N. E., Ryabchikova, T. A.,
Stempels, H. C., Weiss, W. W. 1999, A\&AS, 138, 119

\hangpar LeBlanc, F., Monin, D.  2004, in IAU Symp. 224, The A-Star
Puzzle, ed. J. Zverko, J. \v{Z}i\v{z}novsk\'{y}, S. J. Adelman
\& W. W. Weiss (Cambridge: Cambridge Univ. Press), p. 193


\hangpar Lemke, M. 1997, A\&AS, 122, 285

\hangpar Lucke, P. B. 1978, A\&A, 64, 367

\hangpar Mel\'{e}ndez, M., Bautista, M. A., Badnell, N. R.
2007, A\&A, 469, 1203

\hangpar Michaud, G. 1970, Ap J, 160, 641

\hangpar Moon, T. T. 1984, Comm. Univ. London Obs., No. 78

\hangpar Moon, T. T., Dworetsky, M. M. 1985, MNRAS, 217, 305

\hangpar Munari, U., Zwitter, T. 1997, A\&A, 318, 269

\hangpar Nesvacil, N., Weiss, W. W., Kochukhov, O. 2008, Cont. Astron.
Obs. Ska. Pl., 38, 329

\hangpar N\"{o}rtersh\"{a}user, W., Blaum, K., Icker, P., et al.
1998, Eur. Phys. J. D, 2, 33

\hangpar Paunzen, E., Schnell, A., Maitzen, H. M. 2006, A\&A,
458, 293

\hangpar Preston, G. W. 1974, ARAA, 12, 257

\hangpar Proffitt, C. R., Brage, T., Leckrone, D. S., Wahlgren, G. M.,
et al. 1999, ApJ, 512, 942

\hangpar Rogers, N. Y. 1995, Comm in Asteroseismology, 78

\hangpar Ryabchikova, T. 2008, Cont. Astron. Obs. Sk. Pl., 38, 257

\hangpar Ryabchikova, T. 2005, in Element Stratification in Stars,
40 Years of Atomic Diffusion, ed. G. Alecian, O. Richard,
S. Vauclair, EAS Pub. Ser. 17, 253

\hangpar Ryabchikova, T., Kochukhov, O., Bagnulo, S. 2008, A\&A,
480, 811 (RKB)

\hangpar Ryabchikova, T., Leone, F., Kochukhov, O. 2005, A\&A,
438, 973

\hangpar Ryabchikova, T., Savanov, I. S., Hatzes, A. P.,
Weiss, W. W., Handler, G. 2000, A\&A, 357, 981

\hangpar Ryabchikova, T., Piskunov, N., Kochukhov, O.,
Tsymbal, V., Mittermayer, P., Weiss, W. W. 2002, A\&A,
384, 545 (RPK)

\hangpar Sbordone, L., Bonifacio, P., Castelli, F.,
Kurucz, R. L. 2004, Mem. S. A. It. Suppl., 5, 93.

\hangpar Schlegel, D. J., Finkbeiner, D. P., Davis, M. 1998,
ApJ, 500, 525

\hangpar Sigut, T. A. A. 2001. A\&A, 377, L27

\hangpar Stehl\'{e}, C., Hutcheon, R. 1999, A\&AS, 140, 93

\hangpar Thiam, M., Wade, G. A., LeBlanc, F., Khalack, V. R.
2008, Cont. Astron. Obs. Ska. Pl., 38, 461

\hangpar Wolff, S. C. 1983, The A-Stars: Problems and Perspectives,
NASA-SP 463 (see p. 37).

\hangpar Woolf, V., Lambert, D. L. 1999, ApJ, 521, 414
\label{lastpage}

\end{document}